\documentclass[sort&compress]{jfm}

\usepackage{amsmath,amssymb,mathtools}
\usepackage{graphicx}
\usepackage{dcolumn}
\usepackage{bm}
\usepackage{hyperref}
\usepackage{siunitx}
\usepackage{algorithmic,algorithm}
\usepackage[noabbrev,nameinlink]{cleveref}
\usepackage[labelformat=simple]{subcaption}

\usepackage[normalem]{ulem}
\usepackage{comment}
\usepackage{natbib}
\usepackage{overpic}

\newtheorem{definition}{Definition}
\newtheorem{theorem}{Theorem}

\definecolor{mygray}{gray}{0.8}

\newcommand{\diff}[2]{\frac{\mathrm{d}#1}{\mathrm{d}#2}}
\newcommand{\pdiff}[2]{\frac{\partial#1}{\partial#2}}
\renewcommand{\vec}[1]{\bm{#1}}
\newcommand{\norm}[1]{\left\lVert #1\right\rVert}
\newcommand{\intd}[1]{~\mathrm{d}#1}

\newcommand{\x}{\vec{x}}
\newcommand{\X}{\vec{X}}
\newcommand{\Xast}{\X^{\ast}}

\newcommand{\f}{\vec{f}}
\newcommand{\g}{\vec{g}}
\newcommand{\h}{\vec{h}}
\newcommand{\F}{\vec{F}}
\newcommand{\m}{\vec{m}}

\newcommand{\rproj}[1]{\left\langle#1\right\rangle_{\vec{r}}}

\newcommand{\Mpar}{M_\parallel}
\newcommand{\Npar}{N_\parallel}
\newcommand{\Mperp}{M_\perp}
\newcommand{\Nperp}{N_\perp}
\newcommand{\mpar}{m_\parallel}
\newcommand{\mperp}{m_\perp}

\newcommand{\tensor}[1]{\mathsfbi{#1}}

\newcommand{\ex}{\vec{e}_x}
\newcommand{\ey}{\vec{e}_y}
\newcommand{\ez}{\vec{e}_z}
\newcommand{\gx}{\vec{g}_x}
\newcommand{\gy}{\vec{g}_y}
\newcommand{\gz}{\vec{g}_z}
\newcommand{\lie}[2]{\left[{#1},{#2}\right]}
\newcommand{\R}{\mathbb{R}}
\newcommand{\abs}[1]{\left\lvert{#1}\right\rvert}

\shorttitle{The control of particles in the Stokes limit}
\shortauthor{Walker, Ishimoto, Gaffney, and Moreau}
\title{The control of particles in the Stokes limit}

\author{B. J. Walker\aff{1\corresp{\email{benjamin.walker@maths.ox.ac.uk}}}, K. Ishimoto\aff{2}, E. A. Gaffney\aff{1},
 \and C. Moreau\aff{2}}

\affiliation{\aff{1}Wolfson Centre for Mathematical Biology, Mathematical Institute, University of Oxford, Oxford, OX2 6GG, UK
\aff{2}Research Institute for Mathematical Sciences, Kyoto University, Kyoto, 606-8502, Japan}

\begin{document}
    
\maketitle

\begin{abstract}
There are numerous ways to control objects in the Stokes regime, with microscale examples ranging from the use of optical tweezers to the application of external magnetic fields. In contrast, there are relatively few explorations of theoretical controllability, which investigate whether or not refined and precise control is indeed possible in a given system. In this work, seeking to highlight the utility and broad applicability of such rigorous analysis, we recount and illustrate key concepts of geometric control theory in the context of multiple particles in Stokesian fluids interacting with each other, such that they may be readily and widely applied in this largely unexplored fluid-dynamical setting. Motivated both by experimental and abstract questions of control, we exemplify these techniques by explicit and detailed application to multiple problems concerning the control of particles, such as the motion of tracers in flow and the guidance of one sphere by another. Further, we showcase how this analysis of controllability can directly lead to the construction of schemes for control, in addition to facilitating explorations of mechanical efficiency and contributing to our overall understanding of non-local hydrodynamic interactions in the Stokes limit.
\end{abstract}

\section{Introduction}
\label{sec:intro}

Some notion of control is widely sought in many contexts. In fluid dynamics, particularly on the microscale, experimental techniques such as optical tweezers, flow modulation, and applied magnetic fields are capable of precisely influencing the microfluidic environment, with the potential to realise refined control. The subjects of such control are varied, spanning both synthetic and biological agents, including the canonical spermatozoon \citep{Zaferani2018}. Correspondingly, the intended outcomes of control are broad, from the delivery of therapeutics in medicine \citep{KeiCheang2014,Yasa2018,Tsang2020} to the trapping and sorting of cells in microdevices \citep{Zaferani2018,Walker2018}.

Recently, many studies have sought to control microswimmers. Many examples concern the directed motion of magnetotactic helical microswimmers \citep{Wang2018,Tottori2012,Liu2017,Mahoney2011} part of the broader class of magnetic micromachines \citep{Grosjean2015,Tierno2008,Khalil2020}. Abstracted away from the modality of control, a  noteworthy and increasingly popular approach to the design of control schemes is the application of machine learning techniques\citep{Colabrese2017,Schneider2019}, with a particularly remarkable example being the reinforcement-learning scheme of \citet{Mirzakhanloo2020a} that is used to enact theoretical hydrodynamic cloaking. Many of these modern methodologies are evaluated via a number of simulated or experimental examples, with efficacy justified by successful realisation of control in these cases. Whilst such a trial-based validation is reassuring, there is often an absence of a complimentary theoretical basis that provides rigorous assurances that control is indeed possible from a given configuration, a desirable if not necessary property when seeking to elicit control and guidance in practice. The topic of such an uncommon analysis is that of \emph{controllability}, a broad field that, in this context, seeks to determine the conditions in which a given system can be controlled, querying the theoretical existence of a trajectory in the state space that connect given initial and target configurations. This topic is rich, varied, and often framed abstractly; thus, as a practical introduction, we will aim to concretely summarise elementary but powerful aspects of control theory, with a view to application in the Stokes regime.

Whilst theoretical evaluations of controllability are less common in the microswimming context, there has been extensive exploration of the specific cases or general deformable or propulsive bodies \citep{martin2007control, loheac2013controllability, Loheac2014,Loheac2020} and model microrobots, such as an $N$-link swimmer \citep{alouges2013self, moreau2019local}. These studies have each assessed the controllability of a single object immersed in fluid, though the applicability of their conclusions to multi-swimmer systems remains unclear. Indeed, the long-range hydrodynamic interactions present in the Stokes limit can lead to surprising and complex behaviours, such as flagellar synchronisation \citep{Brumley2014, Bruot2016} and the bound swimming states of algae  \citep{Drescher2009}. With this richness of emergent behaviours, the consideration of multi-particle controllability thereby warrants the treatment of non-local interactions, of particular relevance given recent developments in experimental control via optical tweezers \citep{Zou2020}. One such exploration, though notably in the case of inviscid flows, considered the control of a passive particle in a two-dimensional fluid, with control effected by the movement of a cylinder and the induced flow \citep{Or2009}, though similar such evaluations in Stokes flow is currently lacking, to the best of our knowledge. 

Thus, as a further significant aim of this work, we will investigate the theoretical control and controllability of particles via hydrodynamic interactions, considering both accurate and simplified descriptions of canonical Stokes problems, with such justification currently lacking for even the most simple multi-object settings in the Stokes limit. Further, we will seek to highlight additional benefits of conducting such an analysis, with particular regard to the notion of mechanical efficiency.


Hence, and as the principal aim of this work, we will seek to demonstrate the broad utility of controllability analysis in the context of particle motion and hydrodynamic interactions in the Stokes limit. In doing so, and as an additional aim of this study, we will evaluate and explore the controllability of experimentally motivated and seemingly simple systems of particles in a Stokesian fluid, which, despite their idealised nature, give rise to non-trivial control and hydrodynamic problems. To begin, in \cref{sec:controllability}, we will recount key principles and definitions of control theory, cast in a readily applicable form that may be easily translated to other Stokes problems. Equipped with such a framework, we will then investigate the controllability of multi-sphere systems in detail. In the first instance, motivated by exemplar modalities of actuation on the microscale, in \cref{sec:finite_size}, we will evaluate the degree to which two differently sized spheres can be controlled by the application of forces or torques to one of the particles, then assessing the efficiency of aspects of force-driven control. In doing so, we will make use of both high-accuracy and far-field hydrodynamic approximations to the interactions of these spheres, noting differences in predicted controllability and extensions to many-sphere systems. Next, in \cref{sec:tracer_limit,sec:tracers_flow}, we will consider the same problem setting in the limit where the passive spheres are of vanishing size, moving to establish the controllability of passive tracer particles that are advected by the flow due to a force-driven finite-size sphere, later considering a further simplification of this system. Finally, in \cref{sec:control_schemes}, by highlighting and exemplifying the constructive process of \cite{lafferriere1992differential}, we will demonstrate how these analyses can be used to explicitly formulate controls that effect a desired state change, in this case the targeted motion of spheres along prescribed trajectories by both applied forces and the resulting flows.

\section{Controllability}
\label{sec:controllability}
In this study, we will consider the motion of $N$ spheres of potentially unequal radius in a Newtonian fluid in the regime of vanishing Reynolds number, with the fluid velocity $\vec{u}$ and the pressure $p$ being described by the Stokes equations,
\begin{equation}
    \mu\nabla^2\bm{u}=\nabla p\,.
\label{eq:01-01}
\end{equation}
The flow also satisfies the incompressibility condition $\nabla\cdot\vec{u}=0$ and the fluid viscosity $\mu$ is assumed to be constant. We will assume that all quantities have been appropriately non-dimensionalised, considering dimensionless mobility coefficients throughout. Let $\x_i$ ($i=1,2,\ldots,N$) be the position of the $i$-th sphere, with $\vec{U}_i$ and $\vec{\Omega}_i$ being its translational and rotational velocities, respectively. Here and in \cref{sec:finite_size}, we will focus on the effects of external forces and torques, denoted by $\f_i$ and $\m_i$, respectively, on the $N$-sphere system, motivated by the manipulation of microparticles by optical tweezers and the actuated spinning of magnetic rotors. By the linearity of the Stokes equations, the dynamics of the spheres may be written as \citep{Kim2005}
\begin{equation}
    \begin{bmatrix}
    \vec{U}_1;\cdots;\vec{U}_N;
    \vec{\Omega}_1;\cdots;\vec{\Omega}_N 
    \end{bmatrix}
    =
    \tensor{A}
    \begin{bmatrix}
    \f_1;\cdots;\f_N;
    \m_1;\cdots;\m_N 
    \end{bmatrix}\,.
\label{eq:01-02}
\end{equation}
Here, a semicolon denotes vertical concatenation, with the resulting composite force and velocity vectors being of scalar dimension $6N$. The $6N\times 6N$ tensor $\tensor{A}$, often termed the grand mobility tensor, is symmetric and positive-definite. Further, due to the boundary-value nature of Stokes flow, this mobility tensor is completely determined by the shape of the fluid boundary. Thus, for the systems of spheres considered in this work, this tensor depends only on the positions of the spheres, with individual symmetry naturally removing any rotational dependence. For convenience, let $n=3N$ and $m=6N$, representing the composite dimension of the vectors of sphere positions and velocities, respectively. We introduce the $n$-dimensional position vector
\begin{equation}
\X=[\x_{1};\x_{2};\cdots;\x_{N}]
\label{eq:01-03a}
\end{equation}
and the $m$-dimensional external control vector 
\begin{equation}
\F=[\f_{1};\f_{2};\cdots; \f_{N};\m_{1};\m_{2};\cdots;\m_{N}]\,,
\label{eq:01-03b}
\end{equation}
so that, by linearity, we may write the translational evolution of the sphere system in the form
\begin{equation}
\diff{\X}{t} = \tensor{G}\F\,.
\label{eq:01-04}
\end{equation}
with sphere orientation now implicit. Representing the $n\times m$ tensor $\tensor{G}=\tensor{G}(\X)$ by its column vectors $\g_1,\ldots,\g_m$, with
$\tensor{G}=[\g_1,\g_2,\cdots\g_{m}]$, leads to the system
\begin{equation}
\diff{\X}{t} = \sum_{i=1}^m F_i(t)\g_i (\X)\,,
\label{eq:01-05}
\end{equation}
which is the standard form of a \emph{driftless control-affine system}. Here, the $n$-dimensional vector $\X$ is the \emph{state} of the system, evolving in the \emph{state space} $P \subset \mathbb{R}^n$, and the $F_i(t)$ are our \emph{control functions}, which correspond here to components of applied forces and torques. Of note, the hydrodynamic interactions between the spheres are included in the $m$ column vectors $\g_i$, whose time dependence is only via the state $\X$ due to the Stokes limit, a property that will generally hold for problems of Stokes flow. Following \citet{coron2007control}, to which we refer the interested reader for a detailed summary of the theory of control, we define the term \emph{controllable} (in any time) for the sphere system of \cref{eq:01-05}:

\begin{definition}
The system is \emph{controllable} within a state set $Q \subset P$ if, for any given start state $\X_0 \in Q$, end state $\X_1 \in Q$ and duration of time $T>0$, there exists a control function $\bm{F}(t)$ that transports the set of spheres from $\X_0$ at $t=0$ to $\X_1$ at $t=T$.  
\end{definition}

\begin{figure}
\centering
\begin{overpic}[width=13cm]{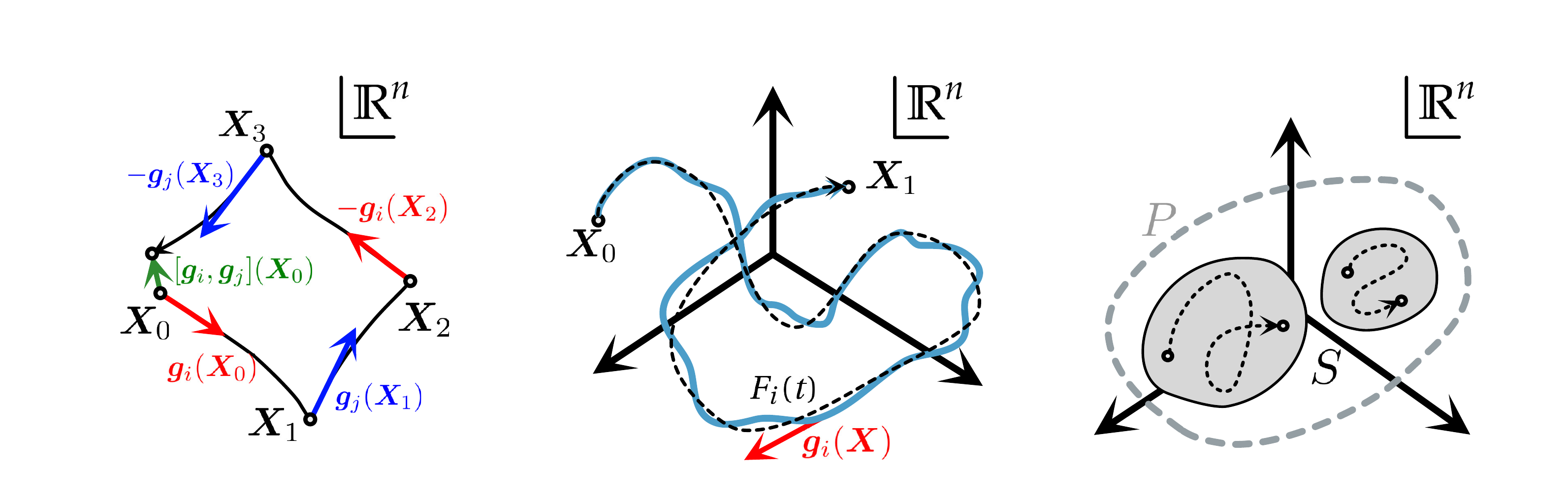}
 \put(5,25){(a)}
 \put(35,25){(b)}
 \put(70,25){(c)}
 \end{overpic}
\caption{Schematic of (a) the Lie bracket, (b) the control system in $\mathbb{R}^n$, and (c) the controllable state set. (a) The Lie bracket, $\lie{\g_i}{\g_j}$ can be computed by considering sequences of controls with infinitesimal duration of time. (b) The controllability of the system guarantees that one can effect evolution from an initial state $\X_0$ to a final state $\X_1$ at a given time $T$. In a driftless control-affine system, the field $\g_i(\X)$ associated with a control $F_i(t)$ is a tangent of the trajectory, but, if the system is controllable, one can nevertheless control the state around a prescribed path (dashed line). The realised path in the state space is shown as a solid curve, approximately coincident with the prescribed path. (c) The controllable state set $S$ is a subspace of $P (\subset \mathbb{R}^n)$, though the system is only controllable within connected subsets of $S$.}
\label{fig:lie}
\end{figure}

Since the field $\g_i$ associated with the control $F_i(t)$ is tangent to the trajectory of the dynamical system, it might seem that the number of scalar controls needs to be larger than the dimension of the system in order to yield a controllable system. However, as two fields do not commute in general, a sequence of controls may generate additional reachable directions and, thus, give rise to controllability in a system of relatively high dimension. This non-commutativity is mathematically represented by a
\emph{Lie bracket}, defined for two fields $\g_i$ and $\g_j$ as 
\begin{equation}
\lie{\g_i}{\g_j}(\X)=
\left(\nabla_{\X}\g_j\right)\g_i-\left(\nabla_{\X}\g_i\right)\g_j\,,
\label{eq:01-11}
\end{equation}
where $\nabla_{\X}\g_j=\left[\pdiff{\g_j}{X_1} \cdots \pdiff{\g_j}{X_n}\right]$ denotes the $n\times n$ Jacobian matrix. The Lie bracket satisfies $\lie{\g_i}{\g_j}=-\lie{\g_j}{\g_i}$, so that $\lie{\g_i}{\g_i}=\vec{0}$, as well as the Jacobi identity.

As an explicit example, for $\tau\ll1$ consider the following sequence of controls on the short time interval $t\in[0, 4\tau]$:
\begin{equation}
\F(t)=\left\{\begin{array}{ll}
F_i=+1,  &\text{ for } t\in[0,\tau)\,, \\
F_j=+1,  &\text{ for } t\in[\tau,2\tau)\,, \\
F_i=-1,  &\text{ for } t\in[2\tau,3\tau)\,, \\
F_j=-1,  &\text{ for } t\in[3\tau,4\tau]\,,
\end{array}\right.
\label{eq:01-12}
\end{equation}
where all unspecified components of $\F$ are assumed to be zero. Under this control, the state $\X(0)$ at $t=0$ evolves to
\begin{equation}
    \X(4\tau)=\X(0)+\tau^2[\g_i, \g_j](\X(0))+O(\tau^3)\,,
\label{eq:01-13}
\end{equation}
as illustrated in \cref{fig:lie}a, with a potential lack of commutativity leading to a non-zero final displacement. This reasoning extends to higher-order Lie brackets, with these terms together spanning all the reachable directions around the state $\X$ via $\F$. This set of all possible Lie brackets, which includes the original fields $\g_i$, is denoted by $\mathrm{Lie}(\g_1,\ldots,\g_m)$, and may be written explicitly as
\begin{equation}
\mathrm{Lie}(\g_1,\ldots,\g_m)=\{ \g_1, \dots,\g_m, [\g_1, \g_2],\ldots,[\g_1, [\g_1, \g_2]],\ldots,[\g_1, [\g_1, [\g_1, \g_2]]]\ldots\}\,.
\label{eq:01-14}
\end{equation}
In what follows, we will refer to the original fields $\g_i$ as the first-order brackets, with $\lie{\g_i}{\g_j}$ being termed the second-order brackets and this convention naturally extending to higher-order terms. The elements of $\mathrm{Lie}(\g_1,\cdots,\g_m)$, each evaluated at a state $\X\in P$, naturally span a vector space $B(\X)$ when seen as elements of $\R^n$. We can then define the \emph{controllable state set} $S\subseteq P$, as illustrated in \cref{fig:lie}c, to be the set of states from which one can move in any direction in the state space, or, perhaps more simply,
\begin{equation}
    S\coloneqq\left\{ \X\in P: \dim{B(\X)}=n\right\} \,.
\label{eq:01-15a}
\end{equation}
As intuition suggests, $S$ is inherently related to the controllability of the system \citep{coron2007control}:

\begin{theorem}[Rashevski-Chow Theorem]
The driftless control-affine system of \cref{eq:01-05} is controllable within any connected subset of the controllable state set $S$.
\end{theorem}

Therefore, in what follows, we will seek to establish the controllability of our $N$-sphere system by determining the dimension of the vector space $B(\X)$ associated with each state $\X\in P$, thereby constructing the controllable state set $S$.

Further, a driftless control-affine system that satisfies this full-rank condition also enjoys a property known as \emph{small-time local controllability} (STLC) at every $\X \in S$, which, in the context of this work, guarantees that a prescribed trajectory in the controllable state space can be followed with arbitrary accuracy, as illustrated in \cref{fig:lie}b. We will revisit this property briefly in \cref{sec:control_schemes} and refer the interested reader to the work of \citet{coron2007control} for a thorough definition of STLC.



\section{Finite size spheres}
\label{sec:finite_size}
\subsection{Geometry and hydrodynamics}
As a particular case of the $N$-sphere system introduced above, we consider the motion of two spheres with the further restriction that only a single sphere may be externally driven. With subscripts of 1 and 2 corresponding to the driven \emph{control sphere} and the other \emph{passive sphere}, respectively, these assumptions amount to taking $\f_2=\m_2=\vec{0}$ and we write $\f_1=\f$ and $\m_1=\m$ for convenience. The dimensionless radii of the two spheres are denoted $a_1$ and $a_2$, respectively, from which we define their ratio $\lambda = a_2/a_1$ and henceforth assume that the radius of the control sphere is unity, without loss of generality. Additionally, we proceed assuming that the spheres do not overlap, so that the state space is given by $P = \{\X\in\R^6: \norm{\x_2-\x_1} > 1 + \lambda\}$.

With these definitions, the general equation of motion \cref{eq:01-04} may be written as
\begin{equation}
\diff{}{t}
\begin{bmatrix}
\x_1 \\ \x_2
\end{bmatrix}
=
\begin{bmatrix}
\tensor{M}\\
\tensor{N}
\end{bmatrix}
\f-
\begin{bmatrix}
\tensor{M}_T\\
\tensor{N}_T
\end{bmatrix}
\m\,,
\label{eq:05-02}
\end{equation}
exploiting the symmetry of two-sphere problem to simplify the form of the hydrodynamic force-velocity relations. Writing $\vec{r}=\x_2-\x_1$ for the position of the passive sphere relative to that of the control sphere, the resulting $3\times3$ tensors $\tensor{M}$, $\tensor{N}$, $\tensor{M}_T$, and $\tensor{N}_T$ are given by
\begin{equation}
\tensor{M}=
\Mpar{}\frac{\vec{r}\vec{r}^T}{\norm{\vec{r}}^2}+\Mperp{}\left(\tensor{I}-\frac{\vec{r}\vec{r}^T}{\norm{\vec{r}}^2}\right)
\text{ and }
\tensor{N}=
\Npar{}\frac{\vec{r}\vec{r}^T}{r^2}+\Nperp{}\left(\tensor{I}-\frac{\vec{r}\vec{r}^T}{\norm{\vec{r}}^2}\right)\,,
\label{eq:05-02b1}
\end{equation}
\begin{equation}
\tensor{M}_T=
M_T\frac{\bm{\epsilon}\cdot\vec{r}}{\norm{\vec{r}}}
\text{ and }
\tensor{N}_T=
N_T\frac{\bm{\epsilon}\cdot\vec{r}}{\norm{\vec{r}}}\,,
\label{eq:05-02b2}
\end{equation}
respectively. Here, $\tensor{I}$ is the $3\times3$ identity tensor and $\bm{\epsilon}$ is the third-rank anti-symmetric tensor. The mobility coefficients $\Mpar$, $\Mperp$, $\Npar$, $\Nperp$, $M_T$, and $N_T$ are all functions of only $r\coloneqq\norm{\vec{r}}$, due to the symmetry of the problem. Further, physical interpretation of each of the coefficients suggests that all but $N_T$ are strictly positive, and that $N_T<0$. However, whilst positivity may be rigorously deduced for $\Mpar$ and $\Mperp$ via calculations of energy dissipation, which we later present in \cref{sec:effi}, we are not aware of similar reasoning that applies to the remaining coefficients. Therefore, to support this intuition, we numerically evaluate these coefficients in \cref{app:mobility}, shown in \cref{fig:mob}, with these findings being consistent with expected signs of these mobility coefficients. Of note, the convergence of the calculation of $M_T$ described in \cref{app:mobility} is not sufficiently fast to conclude that $M_T$ is strictly positive for $\lambda\ll1$, though this will not impact on our later explorations.


If we additionally introduce $\tensor{m} = \tensor{M} - \tensor{N}$ and $\tensor{m}_T = \tensor{M}_T - \tensor{N}_T$, we may further simplify the above into the control-affine form
\begin{equation}
\diff{\tilde{\X}}{t}=\sum_{j} f_j\g_j+\sum_{j} m_j\vec{h}_j\,,
\label{eq:03-03a}
\end{equation}
where $j\in\{x,y,z\}$ and we henceforth adopt these natural Latin subscripts to refer to the components of forces and torques that act in the directions of the right-handed orthonormal triad $\{\ex,\ey,\ez\}$, which forms the fixed basis of the laboratory frame. We similarly index $\g_j$ and $\h_j$, with $\g_j$ and $\h_j$ being the fields associated to $f_j$ and $m_j$, respectively. Here, the modified state and the fields are given by
\begin{equation}
\tilde{\X}=\begin{bmatrix}
\x_1 \\ \vec{r}
\end{bmatrix}\,, \quad
\left[ \g_x,\g_y, \g_z\right]=\begin{bmatrix}
\tensor{M}(r) \\
-\tensor{m}(r)
\end{bmatrix}\,,
\text{ and }
\left[ \h_x,\h_y,\h_z\right]=\begin{bmatrix}
-\tensor{M}_T(r) \\
\tensor{m}_T(r)
\end{bmatrix}\,.
\label{eq:05-02b3}
\end{equation}
More verbosely, defining $\mpar = \Mpar - \Npar$, $\mperp = \Mperp - \Nperp$, and $m_T = M_T - N_T \neq M_T$, the control-affine system of \cref{eq:03-03a} may be written as
\begin{equation}
\diff{\tilde{\X}}{t} = 
\diff{}{t}
\begin{bmatrix}
\x_1 \\ \vec{r}
\end{bmatrix}
=
\begin{bmatrix}
\Mpar{}(r)\frac{\vec{r}\vec{r}^T}{\norm{\vec{r}}^2}+\Mperp{}(r)\left(\tensor{I}-\frac{\vec{r}\vec{r}^T}{\norm{\vec{r}}^2}\right) \\
-\mpar{}(r)\frac{\vec{r}\vec{r}^T}{\norm{\vec{r}}^2}-\mperp{}(r)\left(\tensor{I}-\frac{\vec{r}\vec{r}^T}{\norm{\vec{r}}^2}\right)
\end{bmatrix}
\f +
\begin{bmatrix}
-M_T(r)\frac{\bm{\epsilon}\cdot\vec{r}}{r}\\
m_T(r)\frac{\bm{\epsilon}\cdot\vec{r}}{r}
\end{bmatrix}
\m\,.
\label{eq:05-02c}
\end{equation}
In the remainder of this section, we will consider only this relative two-sphere control system and will, therefore, drop the tilde on the relative composite state vector for notational convenience, hereafter representing the relative state by $\X$.

\begin{figure}
\centering
\includegraphics[width=12cm]{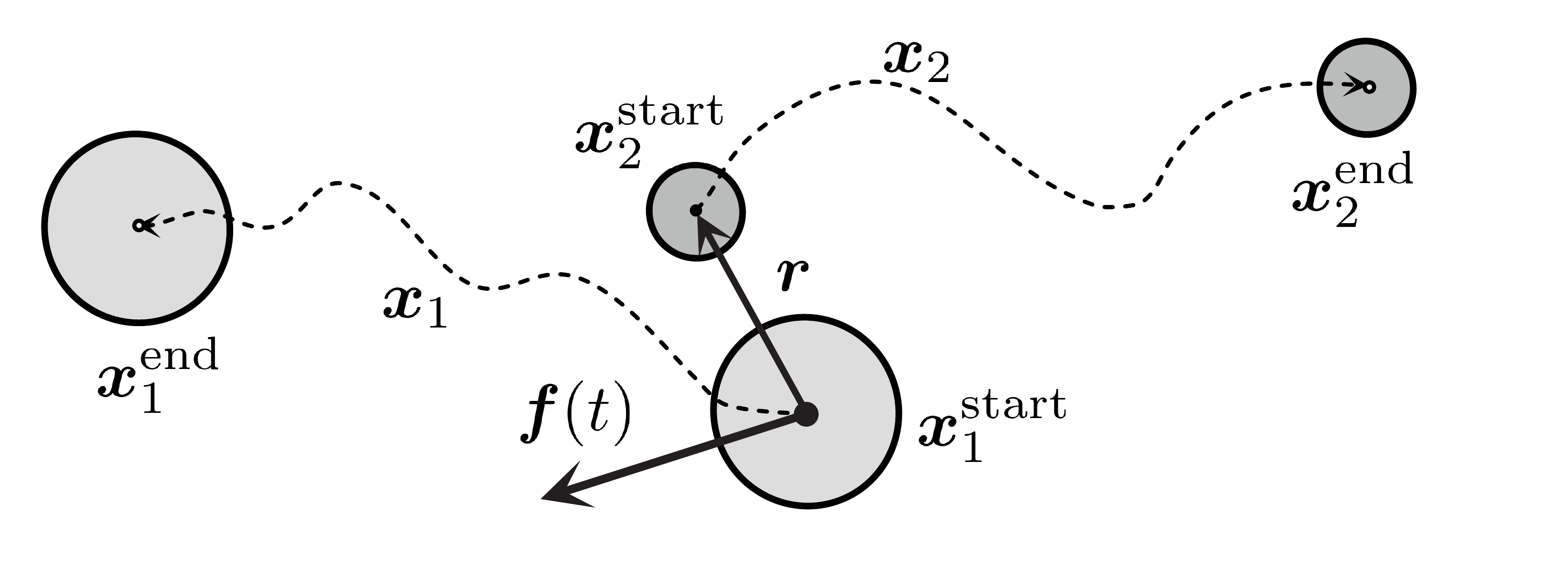}
\caption{Schematic of the control of two spheres by moving one sphere. The force-control problem queries the existence of a forcing function $\f(t)$ that transports the two spheres between given initial and end positions at a given time $T$. Moreover, the full-rank condition in the sphere system allows us to control the spheres to approximately follow prescribed trajectories (dashed lines) in the controllable space.}
\label{fig:conf}
\end{figure}

\subsection{Establishing controllability}
\subsubsection{Control by external force}
\label{sec:control_force}
We first consider the controllability of the two-sphere system when $\m=\vec{0}$, so that only an external force acts on the control sphere through the fields $\g_x$, $\g_y$, and $\g_z$. As we are therefore seeking to control a six-dimensional system with three scalar controls, computation of the Lie brackets of the $\g_i$ is warranted. Writing $\vec{r} = (r_x,r_y,r_z)$ with respect to $\{\ex,\ey,\ez\}$, we compute
\begin{equation}
\lie{\g_i}{\g_j}
=\begin{bmatrix}
\left( \mpar{} M'_\perp-\mperp{} \cfrac{\Mpar{}-\Mperp{}}{r} \right)
\cfrac{r_j\vec{e}_i-r_i\vec{e}_j}{r}
 \\
-\left( \mpar{} m'_\perp-\mperp{} \cfrac{\mpar{}-\mperp{}}{r} \right)
\cfrac{r_j\vec{e}_i-r_i\vec{e}_j}{r}\,,
\end{bmatrix}
\label{eq:04-02}
\end{equation}
where a prime indicates a derivative with respect to $r$. 

The evaluation of the fields $\g_i$ and the second-order brackets $\lie{\g_i}{\g_j}$ at a general state $\vec{X}\in P\subset\R^6$ would yield notationally cumbersome expressions for the Lie bracket, which would be a barrier to further analysis. However, owing to the symmetry of the two-sphere problem, we may, without loss of generality, evaluate the system at a particular state $\Xast{}$, in which the control sphere is located at the origin and the target sphere lies on the $\ex{}$ axis at a distance $r$, i.e. $\x_1 = \vec{0}$ and $\x_2 = r\ex{}$. This reduced configuration is clearly equivalent to a general state up to rotation and translation, with controllability invariant under such transformations in this context. In this parameterised state, the fields $\g_i$ reduce to
\begin{align}
\g_x(\Xast{})&=
\left[
\Mpar{}, 0,  0, -\mpar{}, 0, 0
\right]^T\,, \\
\g_y(\Xast{})&=
\left[
0, \Mperp{},  0, 0, -\mperp{}, 0
\right]^T\,, \\
\g_z(\Xast{})&=
\left[
0, 0, \Mperp{}, 0, 0, -\mperp{}
\right]^T\,.
\label{eq:04-01}
\end{align}
Evaluating the second-order brackets at $\X^{\star}$, we have
\begin{equation}
\lie{\g_x}{\g_y}
(\Xast{})=\begin{bmatrix}
0\\ -\mpar{} M'_\perp+\mperp{} \frac{\Mpar{}-\Mperp{}}{r} \\ 0 \\
0 \\ \mpar{} m'_\perp-\mperp{} \frac{\mpar{}-\mperp{}}{r} \\ 0
\end{bmatrix}\,,
\lie{\g_x}{\g_z}
(\Xast{})=\begin{bmatrix}
0 \\ 0\\ -\mpar{} M'_\perp+\mperp{} \frac{\Mpar{}-\Mperp{}}{r} \\ 
0 \\ 0 \\ \mpar{} m'_\perp-\mperp{} \frac{\mpar{}-\mperp{}}{r} 
\end{bmatrix}\,,
\label{eq:04-03}
\end{equation}
whilst $\lie{\g_y}{\g_z}(\Xast{})=\vec{0}$. Clearly, additional elements are required in order to span a space of dimension six. Hence, we consider the third-order Lie brackets given by
\begin{equation}
\lie{\g_i}{\lie{\g_j}{\g_k}}=\begin{pmatrix}
\vec{L}_{ijk}\\
\vec{\ell}_{ijk}
\end{pmatrix}\,,
\label{eq:04-04b}
\end{equation}
where the three-dimensional field $\vec{L}_{ijk}$ is explicitly given by
\begin{eqnarray}
\vec{L}_{ijk}&=&
\frac{M_{\parallel}-\Mperp{}}{r^3}\left(\mpar{} \mperp{}'-\mperp{}\frac{\mpar{}-\mperp{}}{r}\right)\left( r_jr_k\bm{e}_i+r_j\delta_{ik}\vec{r}-r_ir_k\bm{e}_j-r_i\delta_{jk}\vec{r}\right)\nonumber  \\ 
&& -\frac{\mperp{}}{r} \left(\mpar{} \Mperp{}'-\mperp{}\frac{\Mpar{}-\Mperp{}}{r}\right)
\left[ (\delta_{kj}\bm{e}_i-\delta_{ik}\bm{e}_j) +\left(  \frac{r_ir_k}{r^2}\bm{e}_j-\frac{r_jr_k}{r^2}\bm{e}_i  \right)\right]
 \nonumber \\ 
&&-\frac{\mpar{}}{r^2}\left[\left(\mpar{} \Mperp{}'-\mperp{}\frac{\Mpar{}-\Mperp{}}{r}\right)' \right]\left(r_jr_k\bm{e}_i-r_ir_k\bm{e}_j\right)
\label{eq:04-04c}
\end{eqnarray}
and $\vec{\ell}_{ijk}$ is obtained by replacing $\Mpar{}$ and $\Mperp{}$ with $-\mpar{}$ and $-\mperp{}$, respectively, in the above expression. Here, the Kronecker delta is such that $\delta_{ij}=\vec{e}_i\cdot\vec{e}_j$ and primes again denote differentiation with respect to $r$, with $i,j,k\in\{x,y,z\}$. As before, this cumbersome expression simplifies significantly when evaluating at $\Xast{}$. In particular, $\lie{\g_y}{\lie{\g_x}{\g_y}}$ conditionally generates the additional dimension required to yield controllability and is explicitly given by
\begin{eqnarray}
 &&\lie{\g_y}{\lie{\g_x}{\g_y}}(\Xast{})
\label{eq:04-04} \\
&&=\begin{bmatrix}
-(\Mpar{}-\Mperp{})\left(\cfrac{\mpar{} m'_\perp}{r}-\cfrac{\mperp{}(\mpar{}-\mperp{})}{r^2}\right)
-\mperp{}\left(\cfrac{\mpar{} M'_\perp}{r}-\cfrac{\mperp{}(\Mpar{}-\Mperp{})}{r^2}\right)
 \\ 0 \\0 \\
\mpar{}\left(\cfrac{\mpar{} m'_\perp}{r}-\cfrac{\mperp{}(\mpar{}-\mperp{})}{r^2}\right)
 \\ 0 \\0
\end{bmatrix} \nonumber\,.
\end{eqnarray}
Additional third-order brackets are simply linear combinations of the lower order brackets, with the exception of $\lie{\g_z}{\lie{\g_x}{\g_z}}$, which is parallel to $\lie{\g_y}{\lie{\g_x}{\g_y}}$ by the symmetry of the two-sphere problem. Together, the six terms of \eqref{eq:04-01}, \eqref{eq:04-03} and \eqref{eq:04-04} in general span a space of dimension six, with this property holding unless these fields are linearly dependent. This condition may be succinctly investigated by considering the determinant of the $6\times6$ \emph{controllability matrix}
\begin{equation}
 \tensor{C}=\begin{bmatrix}
 \g_x,  \g_y, \g_z, \lie{\g_x}{\g_y},  \lie{\g_x}{\g_z},  \lie{\g_y}{\lie{\g_x}{\g_y}} 
 \end{bmatrix}\,,
 \label{eq:06-01}
\end{equation}
with the full-rank controllability condition being equivalent to $\tensor{C}$ having a non-vanishing determinant, i.e. $\det{\tensor{C}}\neq0$. This determinant is explicitly given by
\begin{equation}
\det{\tensor{C}} = \frac{\mpar{}^4 \mperp{}^6}{r}\left[ \frac{1}{r}\left( \frac{\Mpar{}}{\mpar{}} -\frac{\Mperp{}}{\mperp{}}\right) - \left(\frac{\Mperp{}}{\mperp{}}\right)' \right]^3\,.
\label{eq:06-02}
\end{equation}

\begin{figure}
\vspace{1em}
\centering
\begin{overpic}[permil,width=0.9\textwidth]{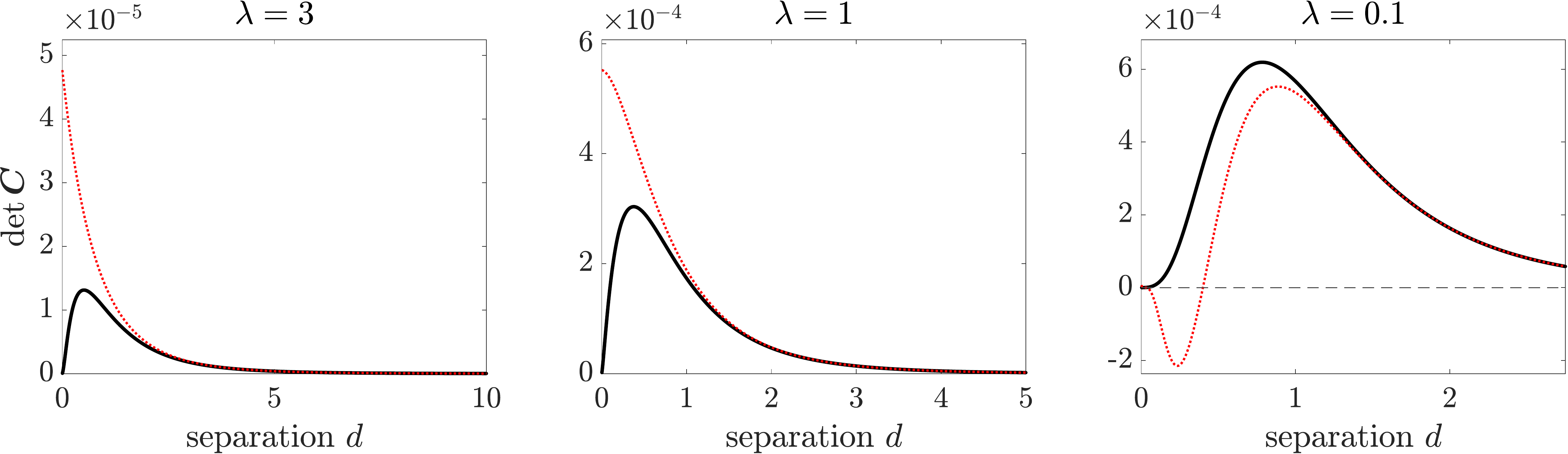}
\put(0,280){(a)}
\put(345,280){(b)}
\put(693,280){(c)}
\end{overpic}
\caption{Values of the determinant $\det{\tensor{C}}$ as a function of separation $d = r - (1+\lambda)$ for different relative sphere radii $\lambda$. The determinants corresponding to the far-field approximation are shown as dotted red curves, whilst those corresponding to the full coefficients are shown in black. The dotted line in (c) corresponds to $\det{\tensor{C}}=0$.}
\label{fig:det}
\end{figure}

In general, closed-form expressions for the mobility coefficients are unavailable, though the coefficients may be obtained as infinite series \citep{Jeffrey1984}. Therefore, as detailed in \cref{app:mobility}, we numerically evaluate these coefficients and compute the value of $\det{\tensor{C}}$, which we show in \cref{fig:det} as solid black lines for a range of separations $d=r - (a_1+a_2) = r - (1+\lambda)$ and relative sphere radii $\lambda$. For $d>0$, we observe that the determinant does not vanish. Thus, we conclude that the controllability matrix $\tensor{C}$ is of rank six throughout the entire admissible state space $P= \{\X\in\R^6: d>0\}$, so that $\dim{\tensor{B}(\X)}=6$ on all of $P$ and, hence, $S = P$. Therefore, by the Rashevski-Chow Theorem, the force-driven two-sphere problem is controllable everywhere, excluding the trivial case when the spheres intersect with one another.

Though we have used accurate expressions for the mobility coefficients in the above analysis, the widespread use of leading-order far-field approximations in the study of motion in Stokes flow motivates repeating the above calculation with the far-field analogues of the mobility coefficients. In this case, having non-dimensionalised forces relative to the mobility coefficient, the far-field approximations to the mobility coefficients are simply
\begin{equation}
\Mpar{} = \Mperp{} = 1\,, \quad \mpar{} = 1- \frac{3}{2r}\,, \quad \mperp{} = 1 - \frac{3}{4r}\,,
\label{eq:06-04a}
\end{equation}
with the far-field approximation to the determinant being
\begin{equation}
\det{\tensor{C}}=\frac{27}{512}\frac{(2r-3)(8r-9)^3}{(4r-3)r^9}\,,
\label{eq:06-04b}
\end{equation}
shown as red dotted lines in \cref{fig:det}. As can be seen 
from this expression, and as is evident in the figure, this approximation does not yield an everywhere-controllable system for $\lambda>1/2$ and this choice of $\tensor{C}$, with the determinant vanishing at $r=3/2$ and $r=9/8$. These zeros can be seen in \cref{fig:det}c and correspond to $d = 1/2 - \lambda$ and $d = 1/8 - \lambda$, respectively. Despite this, as should be expected, good agreement between the two determinant calculations is evident for $d\gtrsim2$, though the far-field theory fails to predict the local maxima in the determinant seen in \cref{fig:det}a and \cref{fig:det}b.

\subsubsection{Control via an external torque}
\label{sec:control_torque}

Complimentary to the above enquiry, we now seek to evaluate the controllability of a torque-driven system, taking $\f=\vec{0}$ instead of $\vec{m}=\vec{0}$. As in the analysis of the force-driven system, we will exploit the symmetry of the two-sphere problem to evaluate $B(\X)$ on a reduced set of states, which we again denote by $\Xast{}$, without loss of generality. First, however, we compute the second-order Lie brackets, which may be written as
\begin{equation}
\lie{\h_i}{\h_j}=
\begin{bmatrix}
\frac{-2m_TM_T}{r^2}\left(r_j\vec{e}_i-r_i\vec{e}_j\right)  \\
\frac{2m_T^2}{r^2}\left(r_j\vec{e}_i-r_i\vec{e}_j\right)
\end{bmatrix}\,,
\label{eq:05-05a}
\end{equation}
where $i,j\in\{x,y,z\}$. Evaluated at a state in the reduced space, these become
\begin{eqnarray}
\lie{\h_x}{\h_y}(\Xast{})&=&
\left[
0, \frac{2m_TM_T}{r}, 0, 0, -\frac{2m_T^2}{r}, 0
\right]^T \\
\lie{\h_x}{\h_z}(\Xast{})&=&
\left[
0, 0, \frac{2m_TM_T}{r}, 0, 0, -\frac{2m_T^2}{r}
\right]^T\,,
\label{eq:05-05}
\end{eqnarray}
and $\lie{\h_y}{\h_z}(\Xast{})=\vec{0}$, whilst the fields corresponding to the control are simply
\begin{eqnarray}
\h_x(\Xast{})&=&
\left[
0, 0,  0, 0, 0, 0
\right]^T \\
\h_y(\Xast{})&=&
\left[
0, 0,  M_T, 0, 0, -m_T
\right]^T \,,\\
\h_z(\Xast{})&=&
\left[
0, -M_T, 0, 0, m_T, 0
\right]^T\,.
\label{eq:05-04}
\end{eqnarray} 
Immediately, and perhaps surprisingly, we see that $\lie{\h_x}{\h_y}$ is parallel to $\h_z$, whilst $\lie{\h_x}{\h_z}$ is parallel to $\h_y$, so that these Lie brackets do not generate additional directions in this example. Thus, so far, we have demonstrated only that $\dim{B(\Xast{})}\geq2$, spanned by $\h_y$ and $\h_z$, noting that $m_T\neq M_T$. In seeking higher-dimensional control, one might be tempted to consider higher-order Lie brackets, such as $\lie{\h_x}{\lie{\h_x}{\h_y}}$.

However, this lower bound on $\dim{B(\Xast{})}$ is in fact sharp, though drawing such a conclusion from direct consideration of $\text{Lie}(\h_x,\h_y,\h_z)$ can be cumbersome in general, in this case requiring inductive arguments on the iterated Lie brackets; hence, we will reason from the dynamical system of \cref{eq:05-02c} . First, it is clear from symmetry that the torque-driven system is not controllable along the $\vec{r}$ direction, with direct calculation of $\mathrm{d}\norm{\vec{r}}^2/\mathrm{d}t$ via \cref{eq:05-02c} highlighting that the distance between the spheres is unchanged throughout the motion. Thus, the dimension of $B(\X)$ is at most five. Further, introducing the weighted midpoint $\x_h=(-N_T\X_0+M_T\vec{r})/m_T$, simple calculation yields that $\mathrm{d}\x_h/\mathrm{d}t=\vec{0}$, irrespective of the applied torque, with the mobility coefficients being explicit functions of only geometry. Hence, with this last constraint imposing three scalar conditions, we can view the system as evolving in a smaller state space $P'$ of dimension 2, namely
\begin{equation}
    P'=\left\{\X\in\mathbb{R}^6:\, r=\textrm{const.}\,,\x_h=\textrm{const.}\right\}\,,
    \label{eq:05-06}
\end{equation}  
and, further, note that the controllable state set within this new state space is found to be $S=P'$.
Therefore, by the Rashevski-Chow Theorem, the torque-driven system is controllable in $S$, indicating that one may control the two spheres around the point $\x_h$ with a fixed distance between them.

\subsection{Efficient control}
\label{sec:effi}
As demonstrated in the above section, control in the entire state space cannot be effected by a simple torque control. Hence, with controllability naturally being a desirable property, we will consider the controllable force-driven problem in further detail. In particular, the presence of local maxima in \cref{fig:det} is suggestive of some notion of optimally efficient control, with the determinant providing an estimate for the volume of the state space explored by the columns of the controllability matrix $\tensor{C}$ (as formally demonstrated in the framework of geometric control theory by the so-called \textit{ball-box theorem} \citep{bellaiche1997tangent}). Motivated by this, we now look to examine the mechanical efficiency of the force-driven control problem of \cref{sec:control_force}.

Let us consider the energy dissipation $E$ that occurs whilst attempting to control the spheres in a particular direction for a small time $\tau$ from a position $\Xast{}$. We will focus on the six fields that make up the columns of $\tensor{C}$ in \cref{eq:06-01}, as they span the space around $\Xast{}$, whilst higher-order brackets are generally associated with lower efficiencies, an example of which we will see later. For a given motion, the energy dissipation rate $\dot{E}$ is given by 
\begin{equation}
\dot{E}=\vec{U}_1\cdot\f_1
\,,\label{eq:07-01a}
\end{equation}
which can be written as the quadratic form
\begin{equation}
\dot{E}(t)= \diff{\X_1}{t}\cdot\f=\f^T\left[ \Mpar{}\frac{\vec{r}\vec{r}^T}{\norm{\vec{r}}^2}+\Mperp{}\left(\tensor{I}-\frac{\vec{r}\vec{r}^T}{\norm{\vec{r}}^2}\right)\right]\f\geq0\,.
\label{eq:07-01b}
\end{equation}
The total energy dissipation $E$ can then be calculated by the time integral of $\dot{E}$.

Let $\gamma$ denote the magnitude of the external force and, in the first instance, set $\f = [\gamma,0,0]^T$. The six-dimensional displacement after time $\tau$ under this control is simply given by $\vec{\Delta}_x = \varepsilon\g_x(\Xast{})$, where we have neglected higher-order terms in $\varepsilon\coloneqq\gamma\tau \ll 1$. Defining the projection map $\rproj{\cdot}$ such that $\rproj{[\vec{a};\vec{b}]} = \vec{b}\in\R^3$ for $[\vec{a};\vec{b}]\in\R^6$, the change in relative position during this motion may be written simply as $\rproj{\vec{\Delta}_x}$. The energy $E_x$ corresponding to this process is given by the integral of \cref{eq:07-01b}, explicitly
\begin{equation}
E_x=\int_0^\tau \gamma^2 \Mpar{}\intd{t}=\gamma^2\tau \Mpar{}=\varepsilon \gamma \Mpar{}
\label{eq:07-01b2}
\end{equation}
to leading order in $\varepsilon$. We now define the mechanical efficiency for this process as the relative change in displacement $\vec{r}$ per unit energy consumption, written
\begin{equation}
\eta_{x}= \frac{\rproj{ \vec{\Delta}_x}\cdot\rproj{\g_x}}{E_x \norm{\rproj{\g_x}}}\,,
\label{eq:07-02}
\end{equation}
where displacement is being measured in the $\g_x$ direction and we are suppressing the argument $\Xast{}$ of $\g_x$ for brevity. Making use of the expression of \cref{eq:04-01}, the unit vector $\rproj{\g_x}/\norm{\rproj{\g_x}}$ is simply equal to $-\ex{}$, though we note that this may not always be the case when considering the far-field approximation of the mobility coefficients.

\begin{figure}
\vspace{1em}
\centering
\begin{overpic}[permil,width=0.9\textwidth]{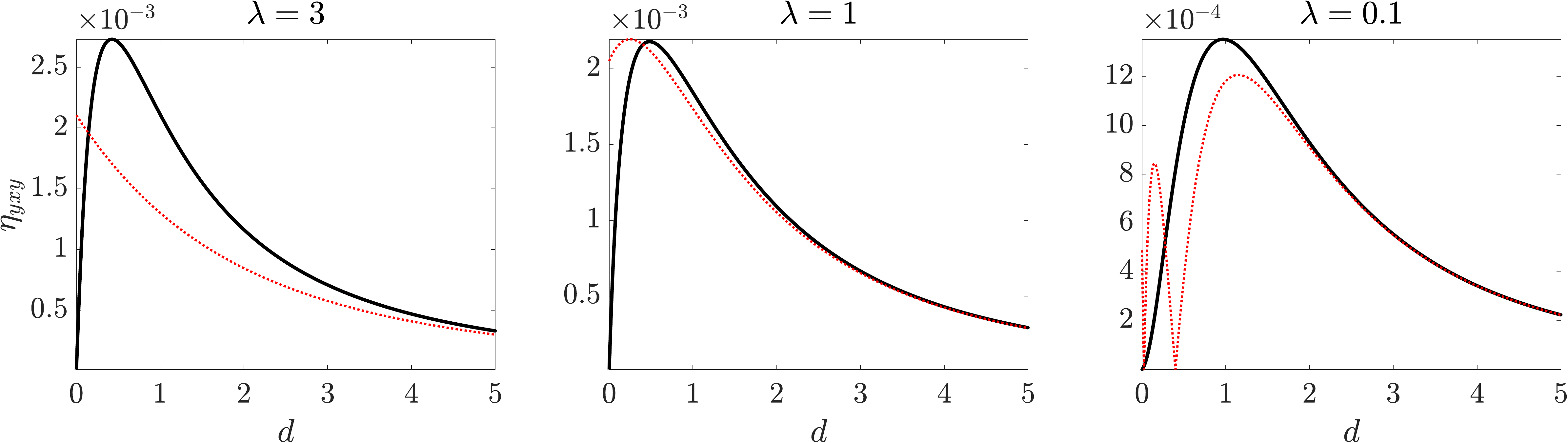}
\put(0,280){(a)}
\put(345,280){(b)}
\put(693,280){(c)}
\end{overpic}
\caption{Values of the mechanical efficiency $\eta_{yxy}$ normalised by $\varepsilon^2/\gamma$ for different sphere radius ratios $\lambda$. The results of the full calculations are shown in black, whilst the analogous results for the far-field hydrodynamic approximation are shown as dotted red curves.}
\label{fig:effi}
\end{figure}

Similarly, we may calculate the displacement and energy consumption for the other control fields $\g_y$ and $\g_z$, giving
\begin{equation}
\vec{\Delta}_y = \varepsilon \g_y(\Xast{})\,,\quad \vec{\Delta}_z =\varepsilon \g_z(\Xast{})\,,\quad E_y=E_z=\varepsilon \gamma \Mperp{}\,.
\label{eq:07-03}
\end{equation}
We may then define $\eta_{y}$ and $\eta_{z}$ as the relative displacement in the $\g_y(\Xast{})$ and $\g_z(\Xast{})$ directions, respectively, per unit mechanical energy consumption by the control, analogously to \cref{eq:07-02}. These definitions can be easily extended to the other elements of $B(\X)$, recalling that Lie brackets represent piecewise sequences of applied external forces. The displacements and energies associated with $\lie{\g_x}{\g_y}$ and $\lie{\g_x}{\g_z}$ are, to leading order in $\varepsilon$,
\begin{equation}
\vec{\Delta}_{xy}=\varepsilon^2 \lie{\g_x}{\g_y}(\Xast{})
\,,\quad \vec{\Delta}_{xz} = \varepsilon^2 \lie{\g_x}{\g_z}(\Xast{})\,,\quad E_{xy}=E_{xz}=2\varepsilon \gamma (\Mpar{} + \Mperp{})\,,
\label{eq:07-04}
\end{equation}
respectively. The iterated Lie bracket $\lie{\g_y}{\lie{\g_x}{\g_y}}$ gives rise to a displacement on the order of $\varepsilon^3$,
\begin{equation}
\vec{\Delta}_{yxy}=\varepsilon^3\lie{\g_y}{\lie{\g_x}{\g_y}}(\Xast{})\,,
\label{eq:07-05}
\end{equation}
with brackets of higher order giving rise to displacements of even higher order. The energy consumption of this term is given by
\begin{equation}
E_{yxy}=2\varepsilon \gamma  (2\Mpar{}+3\Mperp{})\,,
\label{eq:07-06}
\end{equation}
which we note is the same order of $\varepsilon$ as the other Lie brackets.

We now similarly define the mechanical efficiencies of each of the considered Lie brackets, denoting them by $\eta_{xy}$, $\eta_{xz}$, and $\eta_{yxy}$ for each of $\lie{\g_x}{\g_y}$, $\lie{\g_x}{\g_z}$, and $\lie{\g_y}{\lie{\g_x}{\g_y}}$, respectively, the latter of which is given explicitly by
\begin{equation}
\eta_{yxy}=\varepsilon^2\frac{\rproj{\vec{\Delta}_{yxy}}\cdot\rproj{\g_{yxy}}}{\gamma  (4\Mpar{}+6\Mperp{})\norm{\rproj{\g_{yxy}}}}
\label{eq:07-07}\,,
\end{equation}
making use of the compact notation $\g_{yxy}\equiv\lie{\g_y}{\lie{\g_x}{\g_y}}$. In \cref{fig:effi}, we compute $\eta_{yxy}$ for various relative radii $\lambda$, each normalised by $\varepsilon^2/\gamma$. For completeness, we also repeat the above calculations using the far-field approximation for the hydrodynamics, with the resulting efficiencies plotted as dotted red curves in \cref{fig:effi}. Of note, the direction of $\lie{\g_y}{\lie{\g_x}{\g_y}}(\Xast{})$ reverses around $d=0.5$ in \cref{fig:effi}c for the far-field case, giving rise to a non-smooth efficiency. Returning to the full mobility coefficients, we also note the presence of a local maximum in each of these plots, a property shared with $\eta_{xy}$ (not shown), which corresponds to the energetically optimal distance at which this control should be applied. 

\begin{figure}
\begin{center}
\begin{overpic}
[width=6cm]{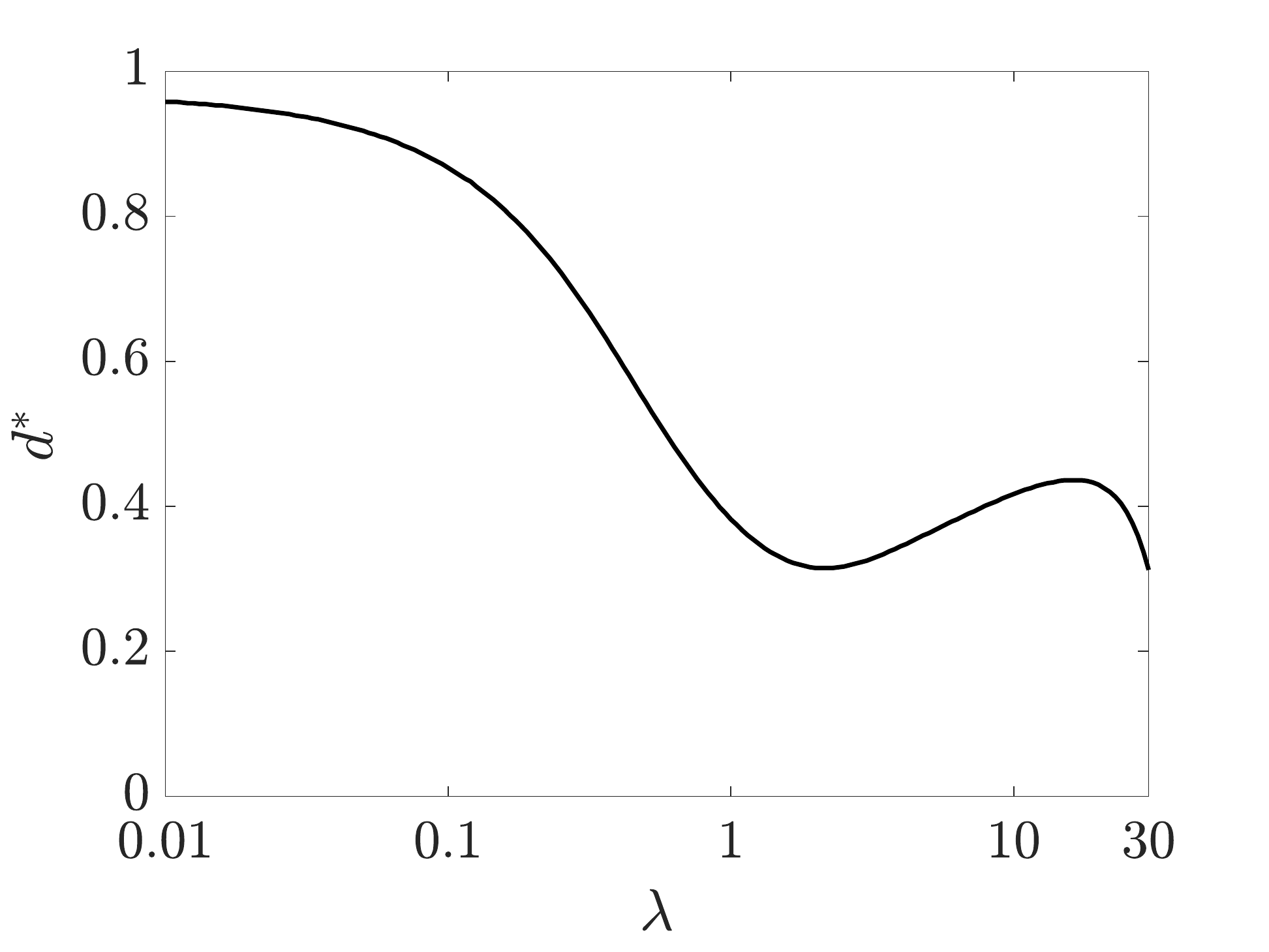}  \put(-10,65){(a)}
\end{overpic}
\begin{overpic}
[width=6cm]{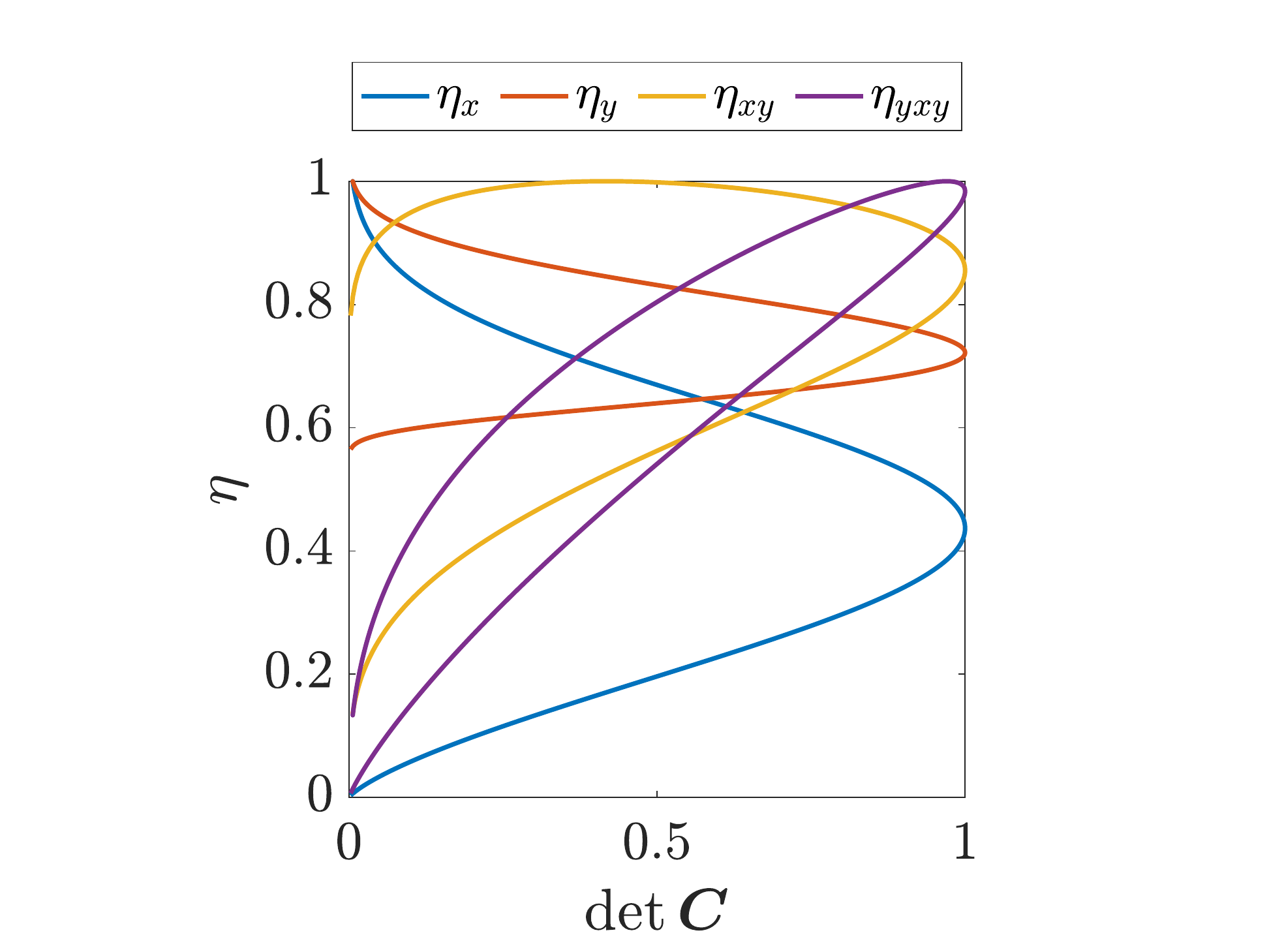}\put(10,65){(b)}
\end{overpic}
\caption{Efficiency, determinant, and optimisation. (a) Optimal distance in terms of the mechanical efficiency $\eta_{yxy}$ as a function of relative sphere radius $\lambda$, with the $\lambda$ axis scaled logarithmically. (b) The relationship between the determinant of the controllability matrix and the mechanical efficiencies, plotted for $\lambda=1$. All quantities in (b) are normalised by their respective maxima.}
\label{fig:optimdist}
\end{center}
\end{figure}

We compute this optimal separation $d^{\ast{}}$ as a function of the relative radius $\lambda$, shown in \cref{fig:optimdist}a, from which we observe that the optimal separation increases to $d^{\ast}\approx1$ as the size of the passive particle decreases ($\lambda\rightarrow0$) and have a minimum around $\lambda=1$, when the spheres are of equal size. Further, as $\eta_{yxy}=O(\varepsilon^2)$, whilst the other efficiencies are $O(1)$ or $O(\varepsilon)$, the optimisation of $\eta_{yxy}$ corresponds to improving the minimally efficient component of a control scheme utilising the considered elements of $\text{Lie}(\g_x,\g_y,\g_z)$, which may be desirable in the context of seeking out efficient schemes for sphere control. One might also attempt to optimise the other efficiencies, for example $\eta_{x}$ and $\eta_{y}$, though, in fact, these do not possess the local maxima of $\eta_{xy}$ and $\eta_{yxy}$ (not shown). Instead, these efficiencies are monotonically increasing functions of the sphere separation $d$, highlighting that simple controls, such as $\vec{f}=[\gamma,0,0]^T$, are relatively inefficient methods for altering the separation $\vec{r}$ of the spheres when they are close to one another, as might be intuitively expected, which warrants the use of more complex Lie bracket controls.

Finally, in order to investigate a potential link between mechanical efficiency and the determinant of \cref{fig:det}, we plot the various efficiencies against the corresponding values of $\det{\tensor{C}}$ in \cref{fig:optimdist}b, fixing $\lambda=1$ and normalising each quantity with respect to its maximum. We observe a surprising correlation between the efficiency $\eta_{yxy}$ and $\det{\tensor{C}}$, with their maxima being approximately coincident, though this agreement is not present to the same extent in the other computed efficiencies. Thus, this partially supports the hypothesised link between efficiency and the determinant of the controllability matrix, though it remains pertinent to consider alternative definitions of efficiency and thoroughly explore the design of efficient controls.

\section{The tracer limit}\label{sec:tracer_limit}
\subsection{Far-field hydrodynamics}\label{sec:tracer_far_field}
A natural limit of the force-controlled finite-size sphere scenario is for $\lambda\ll1$, in which the passive spheres are of negligible size relative to the control sphere. These dynamics are of tracer particles being advected by the flow induced by a translating sphere, which, to a first approximation, is simply that of a Stokeslet. Explicitly, with the position of the control sphere as $\x$ and the relative position of the passive tracers as $\vec{r}_1,\ldots,\vec{r}_{N-1}$, mirroring the previous notation, we have
\begin{equation}
    \diff{\vec{r}_k}{t} = \frac{3}{4}\left(\frac{\tensor{I}}{\norm{\vec{r}_k}} + \frac{\vec{r}_k\vec{r}_k^T}{\norm{\vec{r}_k}^3}\right) \vec{f} - \vec{f}\,,\label{eq:tiny_sphere_far_field_system_expanded}
\end{equation}
for $k=1,\ldots,N-1$ and where the final term arises from subtracting $\dot{\vec{x}}=\vec{f}$ to give the relative velocity of the tracers, noting that the dimensionless mobility coefficient is simply unity in the tracer limit. Written together, we have the $3N$-dimensional control system
\begin{equation}\label{eq:tracer_limit_system}
    \diff{}{t}
\begin{bmatrix}
\vec{r}_1 \\ \vdots \\ \vec{r}_{N-1} \\ \x
\end{bmatrix}
=
\begin{bmatrix}
\frac{3}{4}\left(\frac{\tensor{I}}{\norm{\vec{r}_1}} + \frac{\vec{r}_1\vec{r}_1^T}{\norm{\vec{r}_1}^3}\right) - \tensor{I} \\ 
\vdots \\
\frac{3}{4}\left(\frac{\tensor{I}}{\norm{\vec{r}_{N-1}}} + \frac{\vec{r}_{N-1}\vec{r}_{N-1}^T}{\norm{\vec{r}_{N-1}}^3}\right) - \tensor{I} \\
\tensor{I}
\end{bmatrix}\f = \sum_{i\in\{x,y,z\}} f_i \g_i \,.
\end{equation}
We particularly note that the position of the control sphere can only be controlled directly, i.e. not with any Lie brackets that are second order or higher, as the fields $\g_i$ commute with one another in the $\x$ coordinates. Therefore, the problem of controllability reduces to asking if we can control the relative position of the tracers using only Lie brackets, not using the original fields. For brevity, we define the restricted three-dimensional fields $\tilde{\g}_{i,k}$ as the $\vec{r}_k = (r_{x,k},r_{y,k},r_{z,k})$ components of $\g_i$. Fixing $k\in\{1,\ldots,N-1\}$, straightforward calculation shows that an iterated Lie bracket $\vec{l}$ of order $p$ of the fields $\tilde{\g}_{i,k}$ (in particular, $p=1$ corresponds to the original fields) has components of the form $l_j = L_j r_k^{-5(p-1)-3}$. Here, $L_j$ is a homogeneous polynomial of degree $3(p-1)+2$ in $r_{x,k}$, $r_{y,k}$, and $r_{z,k}$, whose coefficients depend on $(\frac{3}{4}-r_k)$, where $r_k=\norm{\vec{r}_k}$. Moreover, given the polynomials $L_j$ of $\vec{l}$, the components of the bracket $[\tilde{\g}_{i,k},\vec{l}]$ can be obtained recursively via the relation
\begin{equation}
\begin{array}{r l}
r_k^{5p+3} \; [\tilde{\g}_{i,k},\vec{l}]_j = & \displaystyle r_k^2 \left( \frac{3}{4}-r_k \right) \left ( \frac{\partial L_j}{\partial r_{i,k}} r_k^2 - p r_{i,k} L_j \right ) + \\
& \displaystyle + \frac{3}{4} \sum_{q=1}^{3} L_q \left ( 3 r_{i,k} r_{j,k} r_{q,k} - r_k^2 ( \delta_{iq} r_{j,k} + \delta _{jq} r_{i,k} - \delta_{ij} r_{q,k} ) \right ).
\end{array}
\end{equation}
After some simplifications, this gives the order two brackets as
\begin{equation}\label{eq:tracer_second_order}
    \lie{\tilde{\g}_{i,k}}{\tilde{\g}_{j,k}} = \frac{3 (9-8r_k)}{16 r_k^4} ( r_{j,k} \vec{e}_i - r_{i,k} \vec{e}_j).
\end{equation}
The expression of \cref{eq:tracer_second_order} allows us to immediately conclude that the second-order Lie brackets span a space of dimension two, except when $r_k=9/8$, which we recall also prohibited controllability in the far-field case of \cref{sec:control_force}. Here, however, the inclusion of third-order brackets, in particular $\lie{\tilde{\g}_x}{\lie{\tilde{\g}_x}{\tilde{\g}_y}}$ and $\lie{\tilde{\g}_x}{\lie{\tilde{\g}_x}{\tilde{\g}_z}}$, is sufficient to ensure that the Lie brackets indeed span a space of dimension at least two.

In order to make further progress without lengthy calculation, we now restrict to the case with a single tracer, taking $N=2$. Hence, we suppress the second subscript of $k$ in $\tilde{\g}_{i,k}$, with $k$ simply being unity. With this simplification, we may again make use of the state-space reduction of \cref{sec:finite_size}, exploiting the invariance of the controllability of this system to translation and rotation, evaluating the Lie brackets at a reduced configuration in which $\x=\vec{0}$ and $\vec{r}_1\equiv\vec{r} = r\ex$, without loss of generality. In this reduced state, the space spanned by $\lie{\tilde{\g}_x}{\lie{\tilde{\g}_x}{\tilde{\g}_y}}$ and $\lie{\tilde{\g}_x}{\lie{\tilde{\g}_x}{\tilde{\g}_z}}$ is spanned by $\ey$ and $\ez$, with the $\ex$ direction seemingly uncontrollable.

In an attempt to yield controllability in this final direction, we consider the $\ex$ components of further Lie brackets, with these components having been identically zero for those brackets considered thus far. The lowest order brackets with a non-zero $\ex$ component are $\lie{\tilde{\g}_y}{\lie{\tilde{\g}_x}{\tilde{\g}_y}}$ and $\lie{\tilde{\g}_z}{\lie{\tilde{\g}_x}{\tilde{\g}_z}}$, with this shared component being explicitly given as
\begin{equation}
    \frac{3}{32}\frac{(2r-3)(8r-9)}{r^5}\,.
\end{equation}
Vanishing when $r=3/2$ and $r=9/8$, this closely mirrors the structure of \cref{eq:06-04b}, with these values seemingly being an emergent feature of the far-field approximation. From these third-order Lie brackets, we can therefore conclude that the two-particle tracer-limit system is controllable by the Rashevski-Chow Theorem, except when $r=3/2$ or $r=9/8$. Whilst these both correspond to the spheres being in close proximity, with the far-field approximation almost certainly being inappropriate here, it is instructive to further consider the former of these two cases, with controllability able to be readily established at $r=9/8$ via either of the fourth-order Lie brackets $\lie{\tilde{\g}_y}{\lie{\tilde{\g}_x}{\lie{\tilde{\g}_x}{\tilde{\g}_y}}}$ and $\lie{\tilde{\g}_x}{\lie{\tilde{\g}_y}{\lie{\tilde{\g}_x}{\tilde{\g}_y}}}$.

In order to best examine the scenario where $r=3/2$, we consider the evolution of the squared separation distance $\norm{\vec{r}}^2$ in full generality, as in \cref{sec:control_torque}. This is simply given by
\begin{equation}
    \diff{\norm{\vec{r}}^2}{t} = 2\vec{r}^T\diff{\vec{r}}{t} = \frac{3-2\norm{\vec{r}}}{\norm{\vec{r}}}\vec{r}^T\f\,,
\end{equation}
vanishing and independent of the control $\f$ when $\norm{\vec{r}}=3/2$. Hence, if the tracer is located at this distance from the control sphere, the particle will remain trapped at this separation, irrespective of the applied control, thus is not controllable from this configuration. It should be noted, however, that confounding factors in a physical system, such as the presence of noise or other objects in the flow, may act to move the particle away from such a zero-measure configuration, with the particle rendered controllable.

Therefore, in summary, we have found that the force-driven sphere-tracer system, with the flow due to the sphere being approximated with simply a Stokeslet and only one tracer being considered, is controllable unless the object centres are separated by a distance of $3/2$, with the controllable state set formally written as
\begin{equation}
    S = \{ \X\in P : \norm{\vec{r}}\neq 3/2 \}\,.
\end{equation}

\subsection{Including the potential dipole}\label{sec:tracer_exact_hydro}
It is well known that the exact solution for Stokes flow around a translating sphere includes a potential-dipole correction to the Stokeslet term of \cref{eq:tracer_limit_system}, which captures finite-volume effects and satisfies the no-slip condition on the surface of the sphere. Hence, noting the difference in controllability found in \cref{sec:finite_size} when using hydrodynamic representations of differing accuracy, we modify the system of \cref{eq:tracer_limit_system}, again taking $N=2$ for simplicity, to include the weighted potential dipole contribution, which we write as
\begin{equation}
    \diff{\vec{r}}{t} = \frac{3}{4}\left(\frac{\tensor{I}}{\norm{\vec{r}}} + \frac{\vec{r}\vec{r}^T}{\norm{\vec{r}}^3}\right) \vec{f} 
    + 
    \frac{1}{4}\left(\frac{\tensor{I}}{\norm{\vec{r}}^3} - \frac{3\vec{r}\vec{r}^T}{\norm{\vec{r}}^5}\right)\f
    -\vec{f}\,.\label{eq:tracer_limit_exact}
\end{equation}
This can be readily seen to satisfy $\mathrm{d}\vec{r}/\mathrm{d}t=\vec{0}$ when $\norm{\vec{r}}=1$. For completeness, the full control system is now given by
\begin{equation}
        \diff{}{t}
\begin{bmatrix}
\vec{r} \\ \x
\end{bmatrix}
=
\begin{bmatrix}
\frac{3}{4}\left(\frac{\tensor{I}}{\norm{\vec{r}}} + \frac{\vec{r}\vec{r}^T}{\norm{\vec{r}}^3}\right) + 
    \frac{1}{4}\left(\frac{\tensor{I}}{\norm{\vec{r}}^3} - \frac{3\vec{r}\vec{r}^T}{\norm{\vec{r}}^5}\right) - \tensor{I} \\ \tensor{I}
\end{bmatrix}\f = \sum_{i\in\{x,y,z\}} f_i \g_i \,.
\end{equation}
We now seek to repeat the analysis of \cref{sec:tracer_far_field} with this improved representation of the hydrodynamics and by doing so assess the impacts of finite-volume effects on tracer controllability. By the same reasoning as above, we seek to establish the controllability of the tracer using only Lie bracket terms, and write $\tilde{\g}_i$ for the $\vec{r}$ components of $\g_i$. The second-order brackets corresponding to this augmented control system are given by
\begin{equation}\label{eq:accurate_tracer_second_order}
     \lie{\tilde{\g}_i}{\tilde{\g}_j} = \frac{-3(r-1)^2 (8r^3 + 7r^2+6r+3)}{16r^8}(r_j \vec{e}_i - r_i \vec{e}_j)\,.
\end{equation}
As with the far-field approximation, these span a space of dimension two, but now without the condition $r=9/8$. Indeed, the polynomial factor in \cref{eq:accurate_tracer_second_order} is non-zero for $r>1$, unsurprisingly vanishing when the tracer touches the control sphere at $r=1$. When evaluated in the reduced configuration, where $\x=\vec{0}$ and $\vec{r}=r\ex$, without loss of generality, these Lie brackets unconditionally span the $\ey{}\ez{}$ plane from any point in the admissible state space $P\coloneqq\{\X\in\R^6: \norm{\vec{r}}>1\}$. Analogously to \cref{sec:tracer_far_field}, the final direction, $\ex$, is generated by the additional brackets $\lie{\tilde{\g}_y}{\lie{\tilde{\g}_x}{\tilde{\g}_y}}$ and $\lie{\tilde{\g}_z}{\lie{\tilde{\g}_x}{\tilde{\g}_z}}$, with their shared $\ex$ entry being
\begin{equation}
    \frac{3(r - 1)^4 (2 r + 1) (8 r^3 + 7 r^2 + 6 r + 3)}{32 r^{11}}\,.
\end{equation}
However, in contrast to the previous analysis, this component is non-zero for all admissible $r$, vanishing at $r=1$ and two points inside the control sphere, with the direction generated unconditionally. Hence, having considered a tracer that is advected by the exact velocity field around a translating sphere, though still neglecting any disturbance by the tracer particle, we can now conclude that the sphere-tracer system is in fact controllable everywhere, in stark contrast to the partial controllability result obtained when considering only the Stokeslet flow induced by the control sphere. This highlights that degree of some care should be taken in accurately specifying the hydrodynamics that govern the evolution of the control system. Nevertheless, we do note some level of robustness, at least in this case, with the difference in theoretical controllability between the Stokeslet and exact flow representations only being for a single separation, though questions of practical controllability, such as mechanical efficiency, may be differently impacted and warrant future exploration.

\section{Tracers in Stokeslet flow}
\label{sec:tracers_flow}
\subsection{Geometry and hydrodynamics}
In order to facilitate progress in the tracer limit above, we simplified the sphere-tracer system to only a single tracer particle. Seeking to consider the controllability of multiple tracer particles in a shared domain, we now make an alternative simplification that allows for ready evaluation of multi-tracer controllability.

In the far-field limit, with the separation of the spheres large compared to their sizes, all interaction terms between the particles vanish, as can be seen explicitly for the two-sphere case in \cref{eq:06-04a}, where $\Mpar{}$, $\Mperp{}$, $\mpar{}$, and $\mperp{}$ all approach unity. With the grand mobility tensor of \cref{eq:01-02} thereby rendered diagonal, it is clear that the control of one particle by another is no longer plausible. Taking this large-separation limit and removing any driving forces or torques on the particles, a scenario that corresponds to a dilute suspension of passive tracers, we consider a different modality of control, one in which tracer motion is effected by an externally imposed flow. Motivated by the far-field approximation used in \cref{sec:tracer_far_field} and recent experimental trappings of self-propelling microswimmers \citep{Zou2020}, we will investigate the motion of tracers in the flow field produced by a dimensionless Stokeslet of strength $\f(t)$, though other flows may be similarly considered. For each tracer, indexed by $k\in\{1,\ldots,N\}$ and with position $\x_k$, its evolution in the laboratory frame is simply given by
\begin{equation}
    \diff{\x_k}{t} = \left(\frac{\tensor{I}}{\norm{\x_k}} + \frac{\x_k\x_k^T}{\norm{\x_k}^3}\right) \vec{f}\,,\label{eq:single_tracer_expanded}
\end{equation}
where the fixed Stokeslet is located at the origin, without loss of generality. Concatenating the $\x_k$ into $\X\in\R^{3N}$ as in \cref{sec:controllability} and forming composite $\g_x$, $\g_y$, and $\g_z$ in the same manner, these equations of motion combine to yield the $N$-tracer driftless control-affine system
\begin{equation}
    \diff{\X}{t} = \sum_{i\in\{x,y,z\}} f_i \g_i\,.
    \label{eq:tracer_affine_system}
\end{equation}

We now restrict to the case where $N=2$, corresponding to the motion of two inert tracers in Stokeslet flow, and will again make use of compact notation for the left-iterated Lie bracket, writing $\g_{xyz}= \lie{\gx}{\lie{\gy}{\gz}}$ and similarly for other iterated brackets, with $\g_{yxxy} = \lie{\gy}{\lie{\gx}{\lie{\gx}{\gy}}}$, for example. With $i,j\in\{x,y,z\}$, the second-order brackets are explicitly given by
\begin{equation}
    \tilde{\g}_{ij,k} = \frac{3}{\norm{\x_k}^2}\left[(\x_k\cdot\vec{e}_j)\vec{e}_i - (\x_k\cdot\vec{e}_i)\vec{e}_j\right]\,,
\end{equation}
where, for $k\in\{1,2\}$, $\tilde{\g}_{\ast{},k}$ denotes the $\x_k$ components of $\g_{\ast{}}$, analogous to the notation of \cref{sec:tracer_limit}. 

\begin{figure}
    \vspace{1em}
    \centering
    \begin{subfigure}{0.45\textwidth}
    \centering
    \begin{overpic}[width =0.7\textwidth]{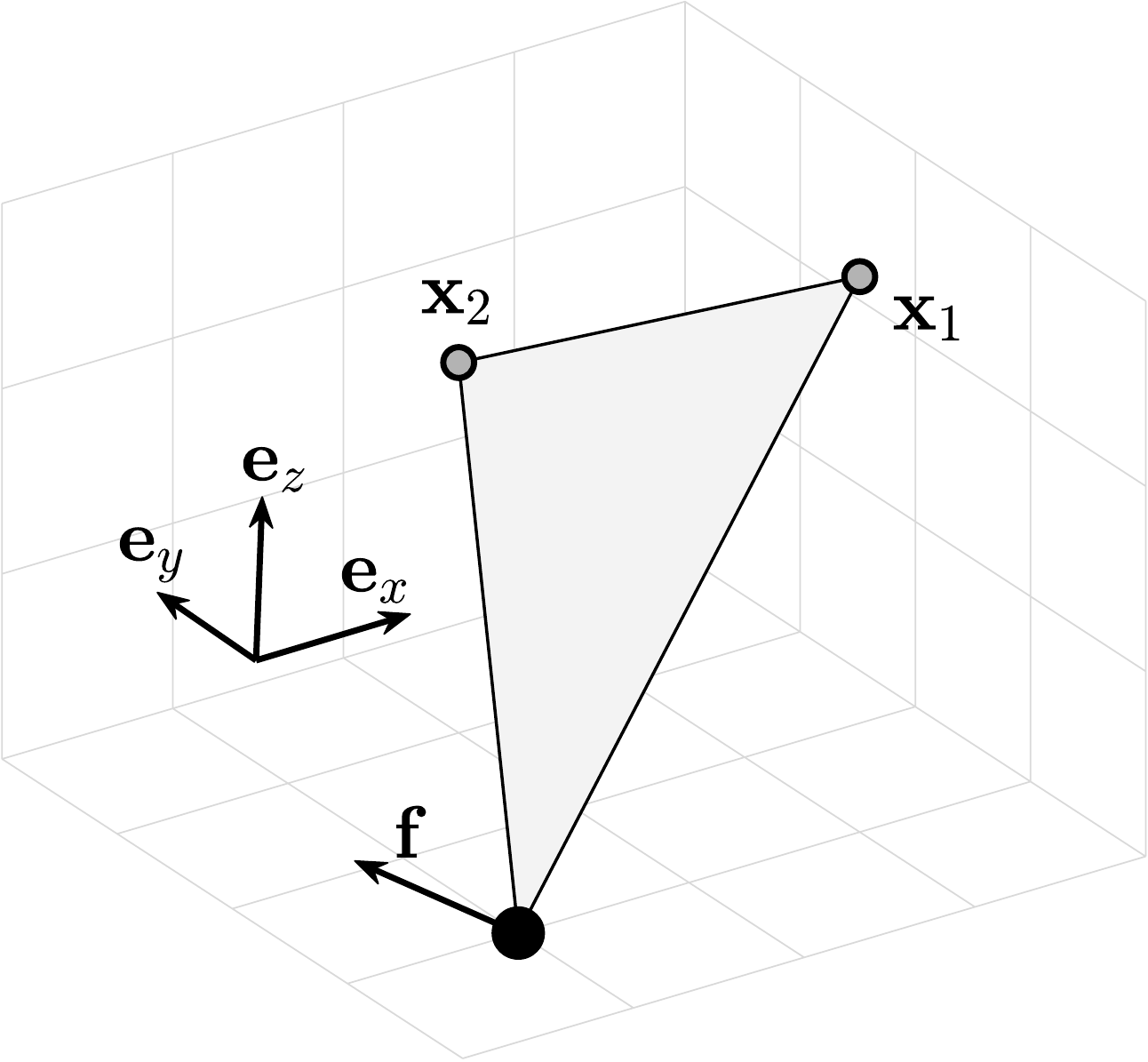}
    \put(-3,90){(a)}
    \end{overpic}
    \end{subfigure}
    \begin{subfigure}{0.45\textwidth}
    \centering
    \begin{overpic}[width =0.8\textwidth]{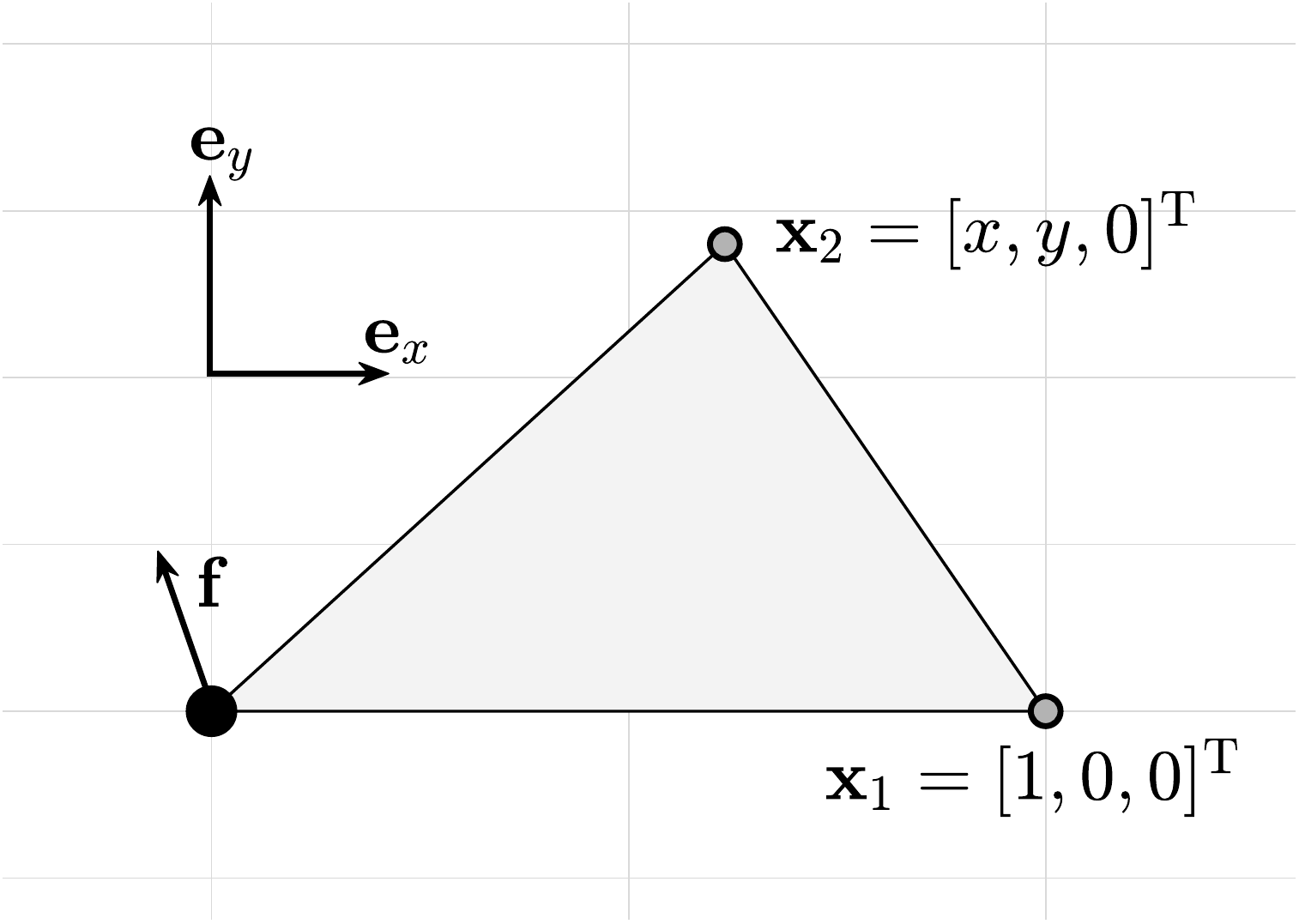}
    \put(-5,72){(b)}
    \end{overpic}
    \end{subfigure}
    \vspace{1em}
    \caption{A reduced two-tracer state space for evaluating controllability. The original six-dimensional state space (a), with $\x_1$ and $\x_2$ general points in $\R^3$ and the Stokeslet situated at the origin, labelled $\f$, is instantaneously equivalent to the reduced planar configuration (b) up to rotation and rescaling, noting the invariance of controllability properties to such transformations in the context of the Stokeslet-driven control system of \cref{eq:tracer_affine_system}. Of note, the reduced system still admits unrestricted motion of each tracer in $\R^3$, with the instantaneous controllability properties merely being evaluated in this notationally reduced but nevertheless general configuration.}
    \label{fig:reduced_two_tracer}
\end{figure}

As in \cref{sec:finite_size}, we will compute the determinants of controllability matrices, seeking to identify the points of the six-dimensional state space $P$ at which these matrices are of full rank, with $P=\{[\x_1;\x_2]\in\R^6 : \x_1,\x_2\neq\vec{0}\}$. However, the evaluation of these expressions would prove to be cumbersome should we retain the six-dimensional representation of the control system. Above, noting the invariance of controllability to rotations and translations in these systems, we successfully reduced this state space to facilitate simple analysis, which remains valid in this case. Indeed, we may now simplify even further, with this tracer system being self-similar with respect to rescalings in space, compensated for by rescalings in time. Thus, we rotate and rescale the spatial description of the two-tracer problem, taking $\x_1=\ex$ and $\x_2=x\ex + y\ey$ and omitting the subscripts on the components of $\x_2$ for notational convenience. We also remark that we may take $y\geq0$, without loss of generality. This significantly reduced configuration is illustrated in \cref{fig:reduced_two_tracer}. 

Of particular note, whilst we will later evaluate controllability matrices at points in this reduced two-dimensional space, with elements denoted by $\Xast{}$, motion out of the plane of this instantaneous configuration is still permitted. Indeed, $\f$ need not lie within this plane. Hence, we will duly seek to identify controllability matrices of rank six, despite evaluating their entries in this notationally and computationally convenient two-dimensional setting. We will refer to this reduced space as $\tilde{P}=\{(x,y)\in\R^2: y\geq0\,,\ (x,y)\neq(0,0)\}$, noting that controllability on $\tilde{P}$ is equivalent to controllability on $P$.

\subsection{Establishing controllability}
\label{sec:two_tracers:controllability}
Seeking a controllability matrix $\tensor{C}$ of rank six, the natural and perhaps minimal choice of $\tensor{C}=[\gx,\gy,\gz,\g_{xy},\g_{xz},\g_{yz}]$ leads to $\det{\tensor{C}}=0$ identically, with higher-order Lie brackets therefore warranted in order to yield controllability. In what follows, we will consider the three controllability matrices
\begin{align}
    \tensor{C}_1 &= [\gx,\gy,\gz,\g_{xy},\g_{yz},\g_{xxy}]\,,\\
    \tensor{C}_2 &= [\gx,\gy,\gz,\g_{xy},\g_{xz},\g_{yxy}]\,,\\
    \tensor{C}_3 &= [\gx,\gy,\gz,\g_{xy},\g_{xz},\g_{yxxy}]\,,
\end{align}
which differ only in the final two columns. For completeness, the full expressions for $\tensor{C}_1$, $\tensor{C}_2$, and $\tensor{C}_3$ are given in \cref{app:two_tracer_analysis} along with their determinants. As also further explained in \cref{app:two_tracer_analysis}, we now identify the controllable sets $S_1,S_2,S_3\subseteq \tilde{P}$ where each of $\tensor{C}_1$, $\tensor{C}_2$, and $\tensor{C}_3$ have non-vanishing determinant, respectively, noting that the system is controllable at a point $\vec{p}\in \tilde{P}$ in the reduced configuration space if $\vec{p}\in S_1\cup S_2\cup S_3$ by the Rashevski-Chow Theorem. 

\begin{figure}
    \centering
    \includegraphics[width=0.7\textwidth]{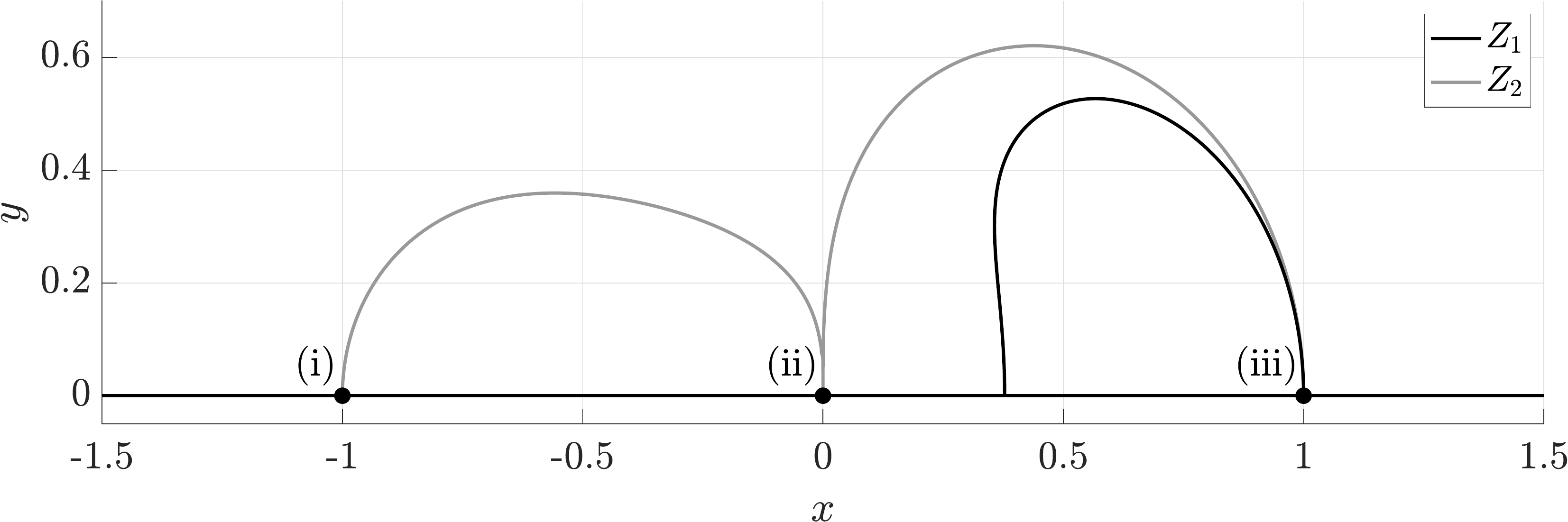}
    \caption{Sets $Z_1$ and $Z_2$ on which $\det{\tensor{C}_1}$ and $\det{\tensor{C}_2}$ in turn vanish, shown as black and grey curves, respectively. Their intersections are marked as black points, corresponding to three cases: (i) $\x_1=-\x_2$, the tracers are mirror images in the Stokeslet; (ii) $\x_2=\vec{0}$, the tracer is at the singular point of the flow, excluded from $P$; (iii) $\x_1=\x_2$, the tracers share location and the system is degenerate.}
    \label{fig:zero_sets}
\end{figure}

The set $Z_1\coloneqq \tilde{P}\setminus S_1$, on which $\det{\tensor{C}_1}$ vanishes, is found to be a subset of the upper right quadrant. The analogously defined $Z_2$ intersects with $Z_1$ at precisely two points $(x,y)\in\{(-1,0),(1,0)\}$, with the origin excluded from $\tilde{P}$. These sets and their points of intersection are shown in \cref{fig:zero_sets}, with controllability established everywhere except these intersections. Notably, the process of identifying potentially uncontrollable states may be streamlined once $Z_1$ is found, with only the intersection $Z_1\cap Z_2$, rather than the whole of $Z_2$, being needed to further interrogate controllability, as demonstrated explicitly in \cref{app:two_tracer_analysis}.

The two points of intersection each correspond to a simple configuration. For $(x,y)=(1,0)$ the tracers overlap and share position for all time, so that $\x_1=\x_2$, with independent control of both $\x_1$ and $\x_2$ clearly impossible. The second case, however, entails only that $\x_1=-\x_2$, so that the two tracers are mirror images with respect to the location of the Stokeslet, a configuration that does not obviously prohibit controllability. Indeed, consideration of the matrix $\tensor{C}_3$ is sufficient to provide controllability in this circumstance, which may be easily established by noting that $\det{\tensor{C}_3}$ is non-zero at $(x,y)=(-1,0)$. Thus, $(-1,0)\notin Z_3$, so that $Z_1\cap Z_2 \cap Z_3 = \{(1,0)\}$. Therefore, $S_1\cup S_2\cup S_3 = \tilde{P}\setminus\{(1,0)\}$ and we have demonstrated controllability on all of the reduced space, excluding the degenerate point where the tracers overlap. Though controllable, the requirement of a fourth-order bracket when $\x_1=-\x_2$ entails that control is somewhat more challenging in this configuration, with higher-order brackets needing more intricate controls to realise in practice.

Recalling that the three considered controllability matrices differed only in their final two columns, a succinct summary of the above calculations is that the composite controllability matrix 
\begin{equation}\label{eq:tracers_flow_overall_cmatrix}
    \tensor{C} =[\gx,\gy,\gz,\g_{xy},\g_{yz},\g_{xz},\g_{xxy},\g_{yxy},\g_{yxxy}]
\end{equation}
is of rank six everywhere in $P$ apart from when the tracers are coincident. Therefore, the two-tracer system is controllable from any such configuration.

\section{Constructing control schemes}
\label{sec:control_schemes}
Controllability analysis theoretically predicts that driving a system from any state $\vec{X}_0$ to any target state $\vec{X}_1$ is possible within the controllable state space, though, as yet, it may be unclear how such a transition can be achieved. In fact, exploiting the directions given by the Lie brackets found to span in the previous sections, it is possible to explicitly design control policies that realise this goal, as described for the first time in the celebrated work of \cite{lafferriere1992differential}, which contributed a general motion planning algorithm for nonholonomic driftless control-affine systems. In this section, we will briefly describe and numerically exemplify this motion planning algorithm as applied to the two-sphere system of \cref{sec:finite_size}. For the sake of legibility, we will reduce the dimension of the system and focus only on planar motion; carrying out the algorithm in higher dimensions is strictly analogous.

Let $X_0$ be the initial state and $X_1$ a target state for the two-sphere system. We have proved in \cref{sec:control_force} that force control of one sphere allows controllability of the system, owing to the controllability matrix composed of four fields, $\tensor{C} = \left ( \g_x, \g_y, \lie{\g_x}{\g_y},\lie{\g_y}{\lie{\g_x}{\g_y}} \right)$, having full rank everywhere in the state space. The principal idea of the motion planning scheme is to utilise these fields and consider the \emph{extended control system}
\begin{equation}
    \dot{\vec{X}} = v_1 \g_x + v_2 \g_y + v_3 \lie{\g_x}{\g_y} + v_4 \lie{\g_y}{\lie{\g_x}{\g_y}} = \tensor{C} \vec{v},
    \label{eq:mp-1} 
\end{equation}
with a new set of four controls $\vec{v} = (v_1,v_2,v_3,v_4)$. Now having as many controls as equations, we can build a control $\vec{v}$ that drives the system along a set trajectory $\vec{\bar{X}}(t)$ that links $\vec{X}_0$ to $\vec{X}_1$ on a set time interval $[0,T]$ by simply inverting system \cref{eq:mp-1}:
\begin{equation}
\vec{v}(t) = \tensor{C}^{-1} \dot{\bar{\vec{X}}}(t).
\label{eq:mp-2}
\end{equation}
We will typically choose $\vec{\bar{X}}$ as a straight line between $\vec{X}_0$ and $\vec{X}_1$, in which case $\vec{\bar{X}}(t) = t \vec{X}_1 + (T-t) \vec{X}_0$ and $\vec{v}(t) = \tensor{C}^{-1} (\vec{X}_1 - \vec{X}_0).$ Note that, for other control systems, such as that of \cref{sec:tracers_flow}, the overall controllability matrix $\tensor{C}$ may not be square, in which case $v$ may be defined with a pseudoinverse of $\tensor{C}$ instead of $\tensor{C}^{-1}$.

By construction, with $\vec{v}$ defined as in \eqref{eq:mp-2}, we have $\vec{X}(T) = \vec{X}_1$. Next, we wish to use this control $\vec{v}$ to build controls $(f_x,f_y)$ that will drive the original two-sphere system to the same target. This can be done by following an elegant, albeit technical, process of calculating so-called \textit{Philip Hall coordinates} of the solution $\vec{X}$, which requires \textit{formal} calculation with noncommutative indeterminates $X$ and $Y$ that represent the fields $\g_x$ and $\g_y$. For more detail on this formalism, we refer the reader to \citet{lafferriere1992differential} or the more comprehensive introduction to geometric control theory of \citet{kawski1997noncommutative}. A Philip Hall basis of the polynomials in $(X,Y)$ up to order 3 is defined as $(X,Y,[X,Y],[Y,[X,Y]],[X,[X,Y]])$, with the first four elements corresponding to the brackets used to define the controllability matrix $\tensor{C}$. Then, one can associated the Philip Hall coordinates $(h_1,h_2,h_3,h_4,h_5)$ to the formal trajectory $\tilde{\vec{X}}$ in such a way that
\begin{equation}
    \tilde{\vec{X}}(t) = e^{h_1(t) X}e^{h_2(t) Y}e^{h_3(t) [X,Y]}e^{h_4(t) [Y,[X,Y]]}e^{h_5(t) [X,[X,Y]]}\,,
    \label{eq:hall}
\end{equation}
decomposing the trajectory with respect to the directions given by each of the (formal) brackets in the Philip Hall basis. Moreover, via the well-known Baker-Campbell-Hausdorff formula, these coordinates $(h_1,h_2,h_3,h_4,h_5)$ may be easily computed as algebraic expressions of the solution to a simple ODE system, which is in turn dependent on $\vec{v}$:
\begin{equation}
    \left \{ 
    \begin{array}{l}
         h_1 = g_1\,, \\
         h_2 = g_2\,, \\
         h_3 = g_3 - g_1 g_2\,, \\
         h_4 = g_4 - g_2 g_3 + \frac{1}{2}g_1 g_2^2 - \frac{1}{6} g_1^2 g_2\,, \\
         h_5 = g_4 - g_1 g_3 + \frac{1}{2}g_1^2 g_2 + \frac{1}{6} g_1 g_2^2\,,
    \end{array}
    \right.
    \quad \text{where} \quad
    \left \{ 
    \begin{array}{l}
         \dot{g}_1 = v_1\,, \\
         \dot{g}_2 = v_2\,, \\
         \dot{g}_3 = v_3 + g_1 v_2\,, \\
         \dot{g}_4 = v_4 + g_1 g_2 v_2\,, \\
         \dot{g}_5 = \frac{1}{2} g_1^2 v_2 + g_1 v_3\,,\\
         g_i(0) =  0\,,\ i = 1\dots 5\,.
    \end{array}
    \right.
    \label{eq:hall-2}
\end{equation}
Finally, the controls $f_x$ and $f_y$ can be used to intuitively generate each term in Eq. \eqref{eq:hall} at $t=T$. For example, the first term $e^{h_1(T) X}$ can be generated by setting $f_x (t) = h_1(T)$ for $t\in[0,1]$. Terms associated with higher-order brackets are obtained through concatenations of $f_x$ and $f_y$, as demonstrated in \cref{eq:01-12} of \cref{sec:controllability}. As an explicit example in this context, the second-order bracket direction $e^{h_3(T) [X,Y]}$ can be achieved on [0,4] with the control policy
\begin{equation}
\vec{f}(t)=\left\{\begin{array}{ll}
f_x=\eta,  &\text{ for } t\in[0,1)\,, \\
f_y=\eta,  &\text{ for } t\in[1,2)\,, \\
f_x=-\eta,  &\text{ for } t\in[2,3)\,, \\
f_y=-\eta,  &\text{ for } t\in[3,4]\,,
\end{array}\right.
\label{eq:hall-3}
\end{equation}
with $\eta = h_3 (T)^{1/2}$ if $h_3(T)\geq0$; if $h_3(T) < 0$, we simply interchange the roles of $f_x$ and $f_y$ in this description. In this example, the controls are specified as piecewise constant functions, but one may instead utilise smoother controls, as further remarked upon in \cref{app:constructing_controls}. Repeating this process for the brackets of third order and performing some minor simplifications, we can define a control policy made of 23 steps that drives $\vec{X}_0$ to $\vec{X}_1$, which we give explicitly in \cref{app:constructing_controls}. However, it should be noted that this control policy only reaches the approximate location of the target, as we have neglected the effects of brackets of order four and higher in the above calculations. To reduce the effects of these higher-order terms, one can subdivide the trajectory into smaller steps by defining intermediate targets, a process with guaranteed convergence \citep{lafferriere1992differential}. In more detail, given a step size of $\Delta$ and a final error tolerance of $\delta$, the algorithm can be summarised as:
\begin{algorithm}
 \begin{algorithmic}
 \STATE $\vec{Y}_0 \leftarrow \vec{X}_0$
 \WHILE{$\| \vec{Y}_0 - \vec{X}_1 \| \geqslant \delta$}
 \STATE $\vec{Y}_{1} \leftarrow \vec{Y}_0 + \min \left ( 1, \frac{\Delta}{\| \vec{X}_0 - \vec{Y}_0 \|} \right ) ( \vec{X}_1 - \vec{Y}_0 )$
 \STATE compute $\vec{v}$ driving $\vec{Y}_0$ to $\vec{Y}_1$ using \eqref{eq:mp-2}
 \STATE compute P. Hall coordinates using \eqref{eq:hall-2}
 \STATE define controls $f_x$, $f_y$ for $t \in [0,24]$ using \eqref{eq:appB-1}
 \STATE compute the trajectory $\vec{Y}$ starting at $\vec{Y}_0$ with controls $f_x$, $f_y$ for $t \in [0,24]$
 \STATE $\vec{Y}_0 \leftarrow \vec{Y}(t=24)$
 \ENDWHILE
 \end{algorithmic}
\end{algorithm}

We numerically apply this algorithm to control our two-sphere system using the MATLAB ordinary differential equation solver \texttt{ode45} \citep{Shampine1997}, with the results displayed in \cref{fig:motion_planning} and Movies 1 and 2 of the Supplementary Material. Panels (a) to (c) and Movie 1 of the Supplementary Material show the character of the trajectory produced by the algorithm on a simple example, with step size $\Delta = 0.5$. The trajectory of each sphere after one iteration is displayed in panel (b), from which we note that generating motion in the target direction requires moving the sphere around a ``loop'' whose length is significantly greater than the final displacement. This serves to validate the study of efficiency in \cref{sec:effi}, which suggested that the third-order brackets require high energy input due the factor $\varepsilon^2$ appearing in \cref{eq:07-06}. With this value of $\Delta$, the algorithm converges after seven iterations, with the results shown in panel (c). The size of the ``loop'' can be reduced by decreasing the value of $\Delta$, which naturally entails that the target will be reached after more iterations. This is shown in panel (e) and Movie 2 of the Supplementary Material, in which the same target is reached with $\Delta = 0.1$ after $27$ iterations. Finally, this algorithm may be applied repeatedly, allowing control of the spheres over more-complex trajectories, as illustrated in panel (f). 

Thus, via this concrete example, we have seen how a controllability analysis may be leveraged practically to design and explicitly construct motion planning policies to realise a control task, with the presented approach also being immediately generalisable. This basic description may be readily augmented and refined, with smoother, more direct trajectories being obtained via a combination of adaptive step sizing and the considered design of control combinations when seeking to generate the Lie bracket directions, of likely pertinence to potential applications and experimental setups.


\begin{figure}
    \centering
    \begin{overpic}[width=\textwidth]{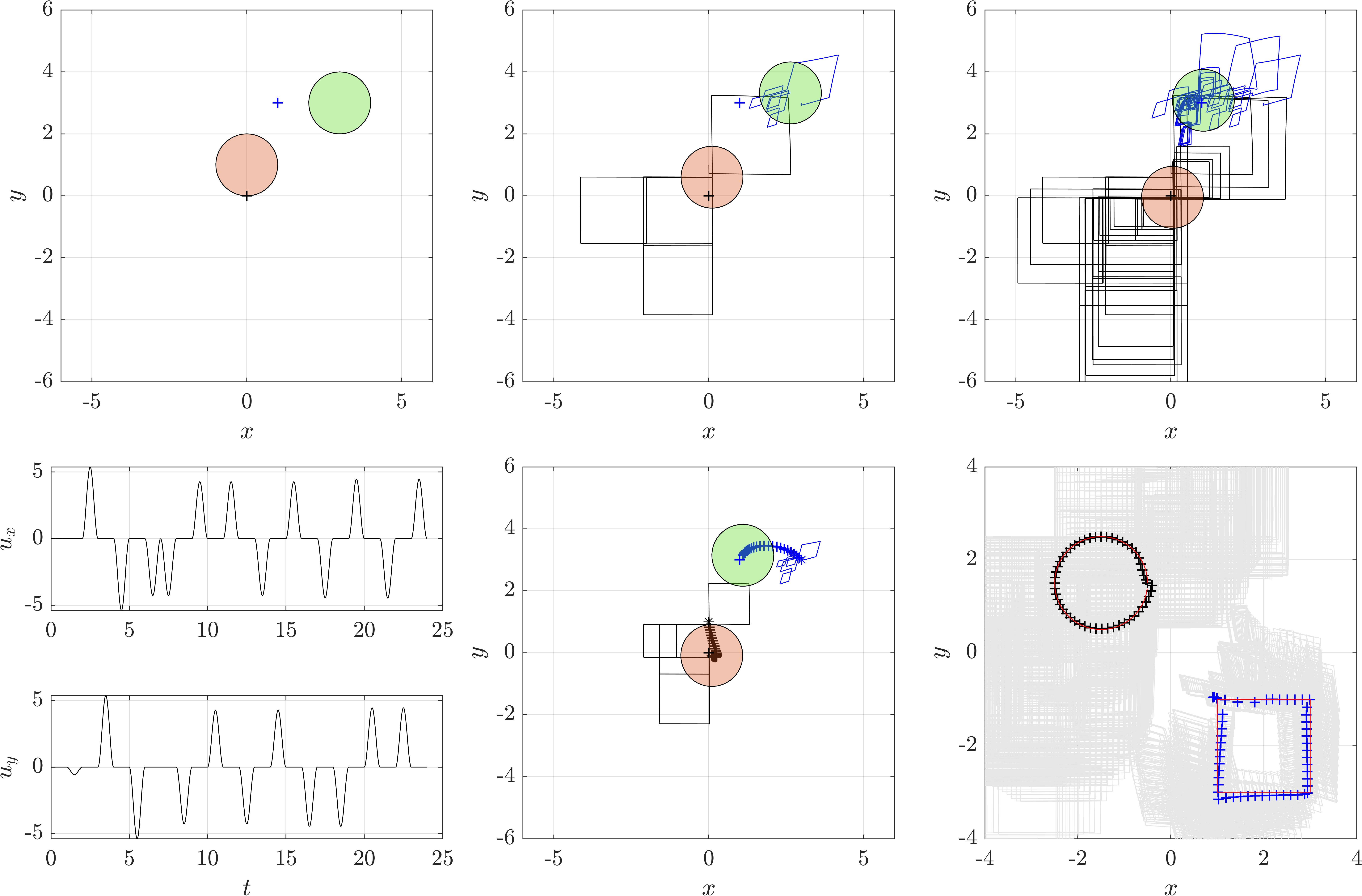}
    \put(0,63){(a)}
    \put(34,63){(b)}
    \put(68,63){(c)}
    \put(0,28.7){(d)}
    \put(34,29){(e)}
    \put(68,29){(f)}
    \end{overpic}
    \caption{Application of the motion planning algorithm to the finite-size two-sphere control problem. The initial configuration is shown on panel (a), with the control sphere in red centred at $(0,1)$ and the passive sphere in green centred at $(3,3)$, both having a radius of unity. The target positions of the spheres, $(0,0)$ and $(1,3)$, are represented on the graph by black and blue crosses, respectively. (b) The position of the spheres and their trajectory after one iteration of the motion planning algorithm, while panel (c) shows the full trajectory that results in the spheres reaching their targets after seven iterations. (d) The controls $f_x$ and $f_y$ to be applied to generate the trajectories on panel (b). (e) The trajectory after one iteration for a smaller value of $\Delta$ ($\Delta = 0.1$), reducing the size of the loops made by the spheres. The black and blue crosses indicate the positions of the spheres after each iteration of the algorithm until they reach the target positions. (f) A more elaborate trajectory where the control and passive spheres are assigned to reach successive targets, along a circle and a square, respectively, each shown as a red line. The full trajectories of the spheres are represented as light grey lines. The respective initial positions of the control and passive sphere are $(-0.5,1.5)$ and $(-1,-1)$ and they traverse their target trajectories anticlockwise and clockwise, respectively. In addition to this figure, Movies 1 and 2 of the Supplementary Material display the trajectories presented on panels (c) and (f), respectively.}
    \label{fig:motion_planning}
\end{figure}
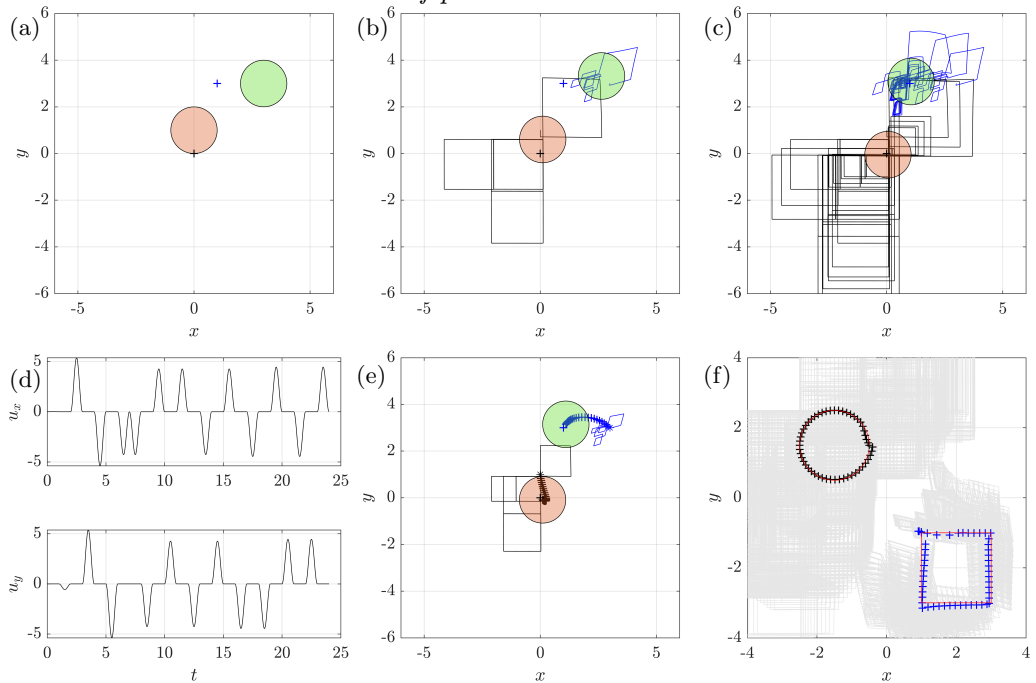

\section{Discussion}
\label{sec:discussion}
In this work, we have presented an introduction to control and controllability in the context of multiple particles in Stokes flow, illustrated by a number of explicit examples of quite general control problems involving hydrodynamic particle-particle interactions, with this framework contrasting with other explorations of controllability in this regime \citep{martin2007control, Loheac2014,alouges2013self,Loheac2020}. In doing so, we have seen how rigorous controllability can be established with elementary, if cumbersome, calculation, highlighting the potential for the systematic justification of controllability in other studies, which is often lacking in contemporary works. This broad applicability draws from the time stationarity of many Stokes flow problems, with explicit reference to time being absent from the governing equations of motion. As we have seen in the cases presented in this study, this can give rise to systems in control-affine form, with simple yet powerful results of geometric control theory then being readily wielded to interrogate the controllability. Absent from the problems considered here, the presence of drift terms, those with no dependence on any control, such as propulsive velocities or prescribed background flows, somewhat complicate the analysis of controllability, though similar tools are able to query the property of STLC, leading to results pertaining to local rather than global control. Such systems are of particular relevance to self-propelled microswimmers, with an analysis of such a swimmer in a controlled background flow being presented by \citet{Moreau2021}.

Though illustrative of a general approach to controllability in the Stokes regime, the particular examples explored in this study may be interpreted in physical or biophysical contexts. Indeed, hydrodynamic interactions between particles have been studied in various biophysical contexts, such as the enhancement of hydrodynamic capturing of swimming 
sperm near an egg \citep{ishimoto2016}, the colonisation of bacteria near a nutrient source \citep{Desai2019}, and bacterial predator-prey dynamics \citep{Ishimoto2020}. Further, flow-mediated particle control has been discussed in the context of nutrition uptake in microorganisms and ciliated cells \citep{Riisgaard2010,Kiorboe2014}. Hydrodynamic interactions between particles are also utilised as a mechanism of cargo transport in an active colloid system \citep{Demirors2018, Wang2019,Yang2020, Zou2020}, with the controllability results established in the present study rigorously guaranteeing that this particle transportation is possible. However, this analysis has also indicated the relative inefficiency of particle repulsion and attraction via hydrodynamic interactions when objects are in very close proximity.

As a consequence of the explicit manner in which we have established control throughout this study, identifying spanning elements of the problem-specific Lie algebra, we have been able to constructively and systematically design schemes for control towards a particular goal, employing the approach of \cite{lafferriere1992differential}. Indeed, by utilising the spanning sets found in our analysis, this methodology constructs an appropriate control, with the resulting control policy demonstrating the key role that non-commutativity can play in realising the manipulation of high-dimensional systems with low-dimensional control modalities. This inherently constructive approach is starkly different to many modern methods for identifying suitable controls, with popular and powerful techniques such as machine learning and high-dimensional nonlinear optimisation being in common use. In contrast to these `black-box' methods, the highlighted explicit approach of \cite{lafferriere1992differential} has the potential to readily develop our understanding of the physics of the control systems considered, with the aforementioned importance of non-commutative controls being a significant example of this. 

Bordering on the related field of optimal control, we have also explored the mechanical efficiency of aspects of control, reporting on the efficiency-optimal distances at which particular controls should be applied. The dependence of these optimal distances on the sphere-size ratio $\lambda$ suggests further investigation in the noted biophysical interpretations of the two-sphere problem, such as the efficient capture of prey and the realisation of sperm-egg encounters \citep{Jabbarzadeh2018}. There also remains significant scope for the consideration of optimality in further generality, with possible extensions of the explicit method of \cite{lafferriere1992differential} being a likely topic of further investigation. In particular, informed by the analysis of the force-driven sphere system in \cref{sec:effi}, we expect that optimal controls for maximising efficiency would make minimal use of terms that correspond to higher-order Lie brackets, with these typically requiring significant additional control input in order to realise comparable changes in state to lower-order brackets. In a practical context, alternative definitions of efficiency, or indeed optimality, may be appropriate, with the investigation of these settings being an additional direction for exploration.

In this work, we have exploited symmetry to improve the computational tractability of the systems that we have considered, a property not present in many-sphere settings or in more complex geometries. Indeed, with confinement having been noted to enhance the controllability of a microrobot \citep{Alouges2013}, assessing the control properties of torque driven spheres near a boundary may be of great interest. With the boundary modifying the mobility coefficients of the spheres, one can imagine that they may be sufficiently modified to enhance controllability, which we have seen is otherwise limited in an unbounded domain. The extension to many-sphere systems, be they finite-size or tracer particles, represents a likely direction for future work; indeed, the principle of the analysis presented in this work is readily applicable to higher-dimensional systems, with the caveat of increased algebraic and notational complexity. However, if such multi-sphere systems are driven with relatively few degrees of freedom, as we have done in this study, their control will almost certainly require the use of high-order Lie brackets, with practical control perhaps limited by this likely necessity.
A further theoretical limitation present in this study lies in the geometric feature of the steady Stokes equations. If the oscillatory frequency of the control force were to be large, the commonplace assumption of negligible inertial effects may no longer be valid. For such a rapidly oscillating force actuation, one might seek to apply the unsteady Stokes equations and the associated Stokeslet, called the oscillet, which leads to a correction in the far-field \citep{Wei2021}.

In summary, we have explored the controllability of multiple physically inspired problems in the Stokes regime, highlighting how explicit assessments of controllability can both facilitate the constructive design of controls and deepen our understanding of underlying fluid mechanics. In examining these example cases, which has also led to an evaluation of mechanical efficiency and its optimisation, we have summarised and demonstrated the application of key principles of geometric control theory such that they may be readily translated to other research contexts, overall serving to illustrate how these simple yet powerful tools can be utilised to rigorously evaluate control in the Stokes limit.

\backsection[Funding]{B.J.W.\ is supported by the UK Engineering and Physical Sciences Research
Council (EPSRC), grant EP/N509711/1. K.I.\ is supported by JSPS-KAKENHI for Young Researchers (18K13456) and JST, PRESTO Grant Number JPMJPR1921. C.M. is a JSPS International Research Fellow (PE20021).}

\backsection[Declaration of Interests]{The authors report no conflict of interest.}

\appendix
\section{Mobility coefficients}
\label{app:mobility}
\begin{figure}
\vspace{1em}
\centering
\includegraphics[width=0.8\textwidth]{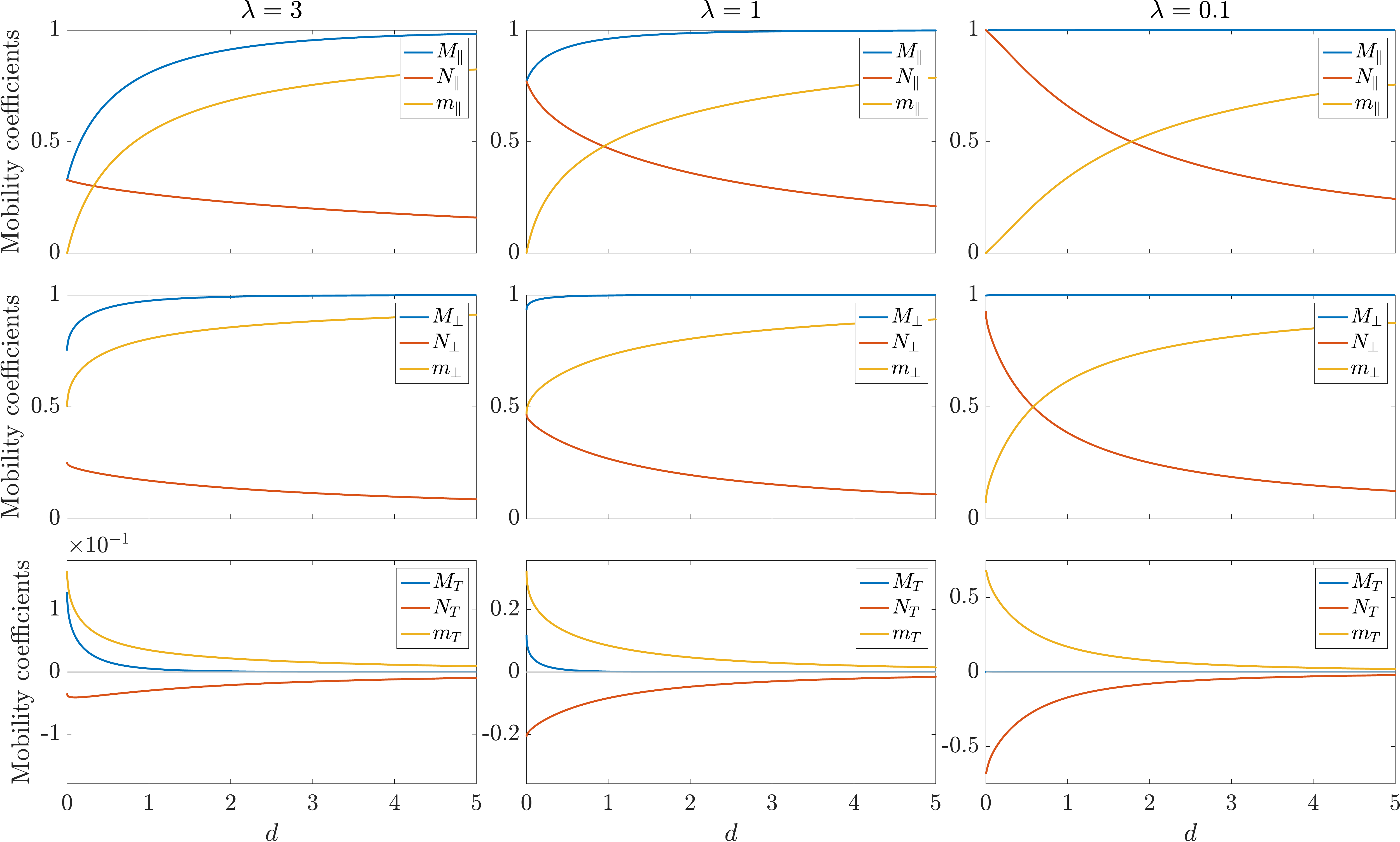}
\caption{Mobility coefficients for various relative sphere sizes, as computed from the expressions of \citet{Jeffrey1984}. For $\lambda=0.1$, $M_T$ is not distinguishable from zero at the resolution of this plot.}
\label{fig:mob}
\end{figure}

The mobility coefficients, $\Mpar{}$, $\Npar{}$,  $\mpar{}$, $\Mperp{}$, $\Nperp{}$, $\mperp{}$, $M_T$, $N_T$, and $m_T$ of \cref{sec:finite_size} are obtained by the inversion of a corresponding resistance matrix. This matrix is constructed from the widely used expressions of \citet{Jeffrey1984}, in which they connect the far-field expansion of the hydrodynamics, up to $O(1/r^{11})$, with results from lubrication theory. These intricate expressions are evaluated numerically, with select cases plotted in \cref{fig:mob}. Of note, for $\lambda\ll1$, the series expression for $M_T$ does not converge sufficiently fast to enable us to conclude that it is non-negative, even if the series is summed up to $O(1/r^{300})$.

\section{Explicit analysis of two tracers in flow}
\label{app:two_tracer_analysis}
With reference to the setup of \cref{sec:two_tracers:controllability}, the controllability matrices $\tensor{C}_1$, $\tensor{C}_2$, and $\tensor{C}_3$ are given explicitly by
\begin{align}
\tensor{C}_1 &= \begin{bmatrix}
    2 & 0 & 0 & 0 & 0 & 0 \\
 0 & 1 & 0 & -3 & 0 & 21 \\
 0 & 0 & 1 & 0 & 0 & 0 \\
 \frac{x^2}{r^{3}}+\frac{1}{r} & \frac{x y}{r^{3}} & 0 & \frac{3 y}{r^{4}} & 0 & -\frac{27 x y}{r^{7}} \\
 \frac{x y}{r^{3}} & \frac{y^2}{r^{3}}+\frac{1}{r} & 0 & -\frac{3 x}{r^{4}} & 0 & \frac{3 \left(7 x^2-2 y^2\right)}{r^{7}} \\
 0 & 0 & \frac{1}{r} & 0 & -\frac{3 y}{r^{4}} & 0
\end{bmatrix} \\
\tensor{C}_2 &= \begin{bmatrix}
 2 & 0 & 0 & 0 & 0 & 6 \\
 0 & 1 & 0 & -3 & 0 & 0 \\
 0 & 0 & 1 & 0 & -3 & 0 \\
 \frac{x^2}{r^{3}}+\frac{1}{r} & \frac{x y}{r^{3}} & 0 & \frac{3 y}{r^{4}} & 0 & \frac{3 \left(2 x^2-7 y^2\right)}{r^{7}} \\
 \frac{x y}{r^{3}} & \frac{y^2}{r^{3}}+\frac{1}{r} & 0 & -\frac{3 x}{r^{4}} & 0 & \frac{27 x y}{r^{7}} \\
 0 & 0 & \frac{1}{r} & 0 & -\frac{3 x}{r^{4}} & 0
\end{bmatrix} \\
\tensor{C}_3 &= \begin{bmatrix}
 2 & 0 & 0 & 0 & 0 & -21 \\
 0 & 1 & 0 & -3 & 0 & 0 \\
 0 & 0 & 1 & 0 & -3 & 0 \\
 \frac{x^2}{r^{3}}+\frac{1}{r} & \frac{x y}{r^{3}} & 0 & \frac{3 y}{r^{4}} & 0 & \frac{27 x y^2 \left(y^2-2 x^2\right)+6 x y^2 \left(x^2-2 y^2\right)-21 x^3 \left(x^2-2 y^2\right)}{r^{12}} \\
 \frac{x y}{r^{3}} & \frac{y^2}{r^{3}}+\frac{1}{r} & 0 & -\frac{3 x}{r^{4}} & 0 & -\frac{3 \left(9 x^2 y \left(x^2+4 y^2\right)+7 x^2 y \left(x^2-2 y^2\right)+2 y^3 \left(2 y^2-x^2\right)\right)}{r^{12}} \\
 0 & 0 & \frac{1}{r} & 0 & -\frac{3 x}{r^{4}} & 0
\end{bmatrix}
\end{align}
where $r=\sqrt{x^2+y^2}$ and these matrices are evaluated at $\Xast{}$ in the reduced two-tracer space $\tilde{P}$. The determinants may be computed as
\begin{align}
    \det{\tensor{C}_1} &= -\frac{108 y^2 \left(7r^4 -8 xr + 1\right)}{r^{13}}\,,\\
    \det{\tensor{C}_2} &= -\frac{108 \left(r^{10} -2 x r^{7} +7 y^2 r^4 + 2 x r^{3}-8 x y^2 r-x^2\right)}{r^{13}}\,,\label{eq:det_M2}
\end{align}
and
\begin{multline}
    \det{\tensor{C}_3} = 54r^{-16} (7 r^{13} -14 x r^{10} +7 r^{9} -7 y^2 r^{7} -7 x r^{6} +14 r^{5} \\+21 x y^2 r^{4} -31 y^2 r^{3} -7 x r^{2}+21 y^4 r+3 x y^2)\,.
\end{multline}

We next identify $Z_1\coloneqq \tilde{P}\setminus S_1$, the set on which $\det{\tensor{C}_1}$ vanishes. This trivially includes all points with $y=0$, with the remaining elements $\Xast{}=[x,y]^T\in\R^2$ necessarily satisfying
\begin{equation}
    8xr = 7r^4+1\,.
\end{equation}
This latter relation admits solutions only when $x>0$ and, thus, defines a curve in the upper right quadrant of the $xy$ plane. With controllability thereby already broadly established, though still lacking on $Z_1$, we now consider the zeros of $\det{\tensor{C}_2}$. Notably, for the purpose of establishing controllability, we may restrict our rootfinding to $Z_1$, computing the intersection $Z_1\cap Z_2$, which significantly simplifies the necessary calculation. First seeking solutions with $y=0$, the condition $\det{\tensor{C}_2}=0$ reduces to 
\begin{equation}
    x^{10} - 2x\abs{x}^7 + 2x\abs{x}^3 - x^2 = 0\,,
\end{equation}
which holds precisely when $x\in\{-1,0,1\}$. Hence, $\{(-1,0),(1,0)\}\subseteq Z_1\cap Z_2$, recalling that $\{(0,0)\}\not\in\tilde{P}$. The remaining elements of $Z_1\cap Z_2$ are solutions of the simultaneous equations 
\begin{align}
    0 & = 7r^4+1 - 8xr\,,\label{eq:zero1}\\
    0 & = r^{10} -2 x r^{7} +7 y^2 r^4 + 2 x r^{3}-8 x y^2 r-x^2\,.\label{eq:zero2}
\end{align}
Recalling that the first equation admits solutions only for $x>0$, this system can be solved analytically via elementary manipulations, leading to the single solution $(x,y) = (1,0)$. Therefore, we in fact have that $Z_1\cap Z_2 = \{(-1,0),(1,0)\}$ precisely, with controllability of the two-tracer system thereby established except at these three points of the reduced space. Both $Z_1$ and $Z_2$ are illustrated in \cref{fig:zero_sets}, computed numerically and with their three points of intersection shown as black dots. The controllability analysis is now readily completed by evaluating $\det{\tensor{C}_3}$ on $Z_1\cap Z_2$, which vanishes only at $(1,0)$. Hence, $S_1\cup S_2\cup S_3 = \tilde{P}\setminus\{(1,0)\}$.

\section{Equations for constructing control schemes}
\label{app:constructing_controls}

The piecewise-constant control policies built from the Philip Hall coordinates $(h_1,h_2,h_3,h_4,h_5)$ are defined on $t=[0,24]$ as follows. We define 
\begin{equation}
    \alpha = h_1(T)\,,\ \beta = h_2(T),\ \gamma = \sqrt{| h_3(T) |}\,,\ \delta = \left ( h_5(T) - \frac{1}{2} \gamma^3 \right ) ^{\frac{1}{3}}\,,\ \epsilon = \left ( h_4(T) + \frac{1}{2} \gamma^3 \right ) ^{\frac{1}{3}}\,.
\end{equation}
The terms $\frac{1}{2} \gamma^3$ in the definition of $\delta$ and $\epsilon$ are used to remove the third-order residuals generated by the control policy generating the direction $[\g_x,\g_y]$. Then, if $h_3(T) \geqslant 0$, we define
\begin{equation}
    f_x(t) = \left \{ \!
    \begin{array}{r l}
         \alpha & \text{on}\;  [0,1), \\
         \gamma & \text{on}\; [2,3), \\
         -\gamma & \text{on}\; [4,5), \\
         \delta & \text{on}\; [6,8) \cup [13,14), \\
         -\delta & \text{on}\; [9,10) \cup [11,12) \cup [15,16), \\
         \epsilon & \text{on}\; [17,18) \cup [21,22), \\
         -\epsilon & \text{on}\; [19,20) \cup [23,24),
    \end{array}
    \right.
    f_y(t) = \left \{ \!
    \begin{array}{r l }
         \beta & \text{on}\; [1,2), \\
         \gamma & \text{on}\; [3,4), \\
         -\gamma & \text{on}\; [5,6), \\
         \delta & \text{on}\; [8,9) \cup [12,13), \\
         -\delta & \text{on}\; [10,11) \cup [14,15),\\
         \epsilon & \text{on}\; [16,17) \cup [18,19), \\
         -\epsilon & \text{on}\; [20,21) \cup [22,23),
    \end{array}
    \right.
    \label{eq:appB-1}
\end{equation}
and, if $h_3(T) < 0$, we instead define
\begin{equation}
    f_x(t) = \left \{ \!
    \begin{array}{r l}
         \alpha & \text{on}\;  [0,1), \\
         \gamma & \text{on}\; [3,4), \\
         -\gamma & \text{on}\; [5,6), \\
         \delta & \text{on}\; [8,9) \cup [12,13), \\
         -\delta & \text{on}\; [10,11) \cup [14,15),\\
         \epsilon & \text{on}\; [16,17) \cup [18,19), \\
         -\epsilon & \text{on}\; [20,21) \cup [22,23),
    \end{array}
    \right.
    f_y(t) = \left \{ \!
    \begin{array}{r l }
         \beta & \text{on}\; [1,2), \\
         \gamma & \text{on}\; [2,3), \\
         -\gamma & \text{on}\; [4,5), \\
         \delta & \text{on}\; [6,8) \cup [13,14), \\
         -\delta & \text{on}\; [9,10) \cup [11,12) \cup [15,16), \\
         \epsilon & \text{on}\; [17,18) \cup [21,22), \\
         -\epsilon & \text{on}\; [19,20) \cup [23,24),
    \end{array}
    \right.
\end{equation}
Smoother controls can be obtained by multiplying the constant value set on an interval $[T_0,T_0+1]$ by any function $\varphi$ supported on this interval whose integral is unity. For the control policies displayed in \cref{fig:motion_planning}d, we have made use of $\varphi(t) = 2 \sin^2 (\pi (t-T_0))$.

\bibliographystyle{jfm.bst}
\bibliography{library.bib}

\end{document}